\documentclass[reprint, amsmath, amssymb,onecolumn, prb]{revtex4-1}
\pdfoutput=1 

\usepackage{graphicx}
\usepackage{dcolumn}
\usepackage{bm}
\usepackage{color}

\begin{document}

\preprint{APS/123-QED}

\title{Order by disorder: saving collective motion\\ from topological defects in a conservative model.} 

\author{Mathias Casiulis}
\affiliation{Sorbonne Universit\'{e}, CNRS, Laboratoire de Physique Th\'{e}orique de la Mati\`{e}re Condens\'{e}e, LPTMC, 4 place Jussieu, Couloir 12-13, 5\`{e}me \'{e}tage, F-75005 Paris, France}
\email{casiulis@lptmc.jussieu.fr}
\author{Marco Tarzia}
\affiliation{Sorbonne Universit\'{e}, CNRS, Laboratoire de Physique Th\'{e}orique de la Mati\`{e}re Condens\'{e}e, LPTMC, 4 place Jussieu, Couloir 12-13, 5\`{e}me \'{e}tage, F-75005 Paris, France}
\author{Leticia F. Cugliandolo}
\affiliation{Sorbonne Universit\'{e}, CNRS, Laboratoire de Physique Th\'{e}orique et Hautes \'{E}nergies, LPTHE, 4 place Jussieu, Couloir 13-14, 5\`{e}me \'{e}tage, F-75005 Paris, France}
\author{Olivier Dauchot}
\affiliation{UMR Gulliver 7083 CNRS, ESPCI Paris, PSL Research University, 10 rue Vauquelin, 75005 Paris, France}

\date{\today}

\begin{abstract}
Using analytic and numerical methods, we study a $2d$ Hamiltonian model of interacting particles carrying ferro-magnetically coupled continuous spins which are also locally coupled to their own velocities. 
This model has been characterised at the mean field level in a parent paper. 
Here, we first obtain its finite size ground states, as a function of the spin-velocity coupling intensity and system size, with numerical techniques.
These ground states, namely a collectively moving polar state of aligned spins, and two non moving states embedded with topological defects, are recovered from the analysis of the continuum limit theory and simple energetic arguments that allow us to predict their domains of existence in the space of control parameters. 
Next, the finite temperature regime is investigated numerically. 
In some specific range of the control parameters, the magnetisation presents a maximum at a finite temperature.
This peculiar behaviour, akin to an order-by-disorder transition, is explained by the examination of the free energy of the system and the metastability of the states of minimal energy. 
The robustness of our results is checked against the geometry of the boundary conditions and the dimensionality of space.
\end{abstract}

\maketitle

\section{Introduction}
Collective motion, the macroscopic ordering of velocities in a many-body system, is a key feature of assemblies of self-propelled particles. 
This phenomenon has recently drawn a lot of attention from the statistical physics community~\cite{Ramaswamy2010,Marchetti2013}, and has been reported and extensively studied in a variety of systems, both natural, like flocks of birds~\cite{Cavagna2010} or bacterial colonies,~\cite{Zhang2010} and artificial, like active colloids~\cite{Bricard2013} or vibrated grains~\cite{Deseigne2010}.
Interacting systems of self-propelled particles have also been the focus of many theoretical studies, starting with the celebrated Vicsek model~\cite{Csahok1995}, and branching into many more recent analytical and numerical works~\cite{Bechinger2016}. 

In typical Vicsek-like models moving individuals are represented as polar self-propelled particles, i.e., particles set in motion by a driving force directed along the heading vector of the particle. 
Assuming, that the speed is always constant leads to minimal spin models where particles  move along the direction of their spin and interact with their neighbours so as to align their velocity vectors, giving rise to collective motion.
An important simplification of Vicsek-type models is to identify the direction of motion with that of the spin of the particles. 
However, as soon as the particles experience other types of interactions, as simple as rigid body repulsion, it is a priori relevant to consider the spins and velocities as distinct dynamical variables. 
It can also be the case that the dynamics of the spin be coupled to that of the velocity, in such a way that, not only the particle aligns its velocity on its spin, but also it aligns its spins on its velocity. 
This has been exemplified as the cause for collective motion in the case of self-propelled granular disks~\cite{Weber2013,Lam2015}, for which the steric interaction alone does not ensure alignment of the velocities, and more recently in the effective dynamics of topological defects in active nematics~\cite{Shankar2018}. 
It was also demonstrated as being a necessary condition for describing the dynamics of a motorised polar particle in a harmonic potential~\cite{Dauchot2019}.

Coupling spin to velocity is expected to induce new physics, not only in the realm of active matter, but also in the context of usual Hamiltonian dynamics. 
Such couplings break Galilean invariance. Broken Galilean invariance is also a hallmark of active systems. 
One may thus wonder if some of the remarkable features of active matter could also occur in conservative systems.
Such questions have been recently discussed in the context of a conservative two-dimensional particle model~\cite{Bore2016}, in which particles carry a continuous and classical spin. 
The model, defined in Eqs.~(\ref{eq:Lagrangian}) and~(\ref{eq:Ham}), is characterised by ferromagnetic interactions between spins of two different particles, and a ferromagnetic coupling between the spin and velocity of the same particle. Because of this coupling, not only Galilean invariance is broken, but also the conserved linear momentum associated to translation invariance is not proportional to the velocity of the centre of mass and the dynamics are not invariant under a global rotation of the spins alone. 
This, in principle, leaves room for collective motion.
A preliminary study of the model has shown that a transition to collective motion does indeed take place  in the mean-field (fully-connected) limit, in spite of momentum conservation~\cite{Bore2016}.

In the present paper, using molecular dynamics simulations and analytic calculations, we characterise the $2d$ behaviour of this model in states such that the system is homogeneous, focusing on the effects of the size, dimensionality, temperature, and spin-velocity coupling intensity.
We report a finite-size magnetisation crossover that is accompanied by a collective ordering of the velocities at low enough temperatures and low enough values of the spin-velocity coupling constant, hereafter defined as $K$.
This ordering breaks down at a finite value of $K$, that depends on the size of the system, as accurately predicted using simple energetic arguments.
The resulting non-magnetised ground states feature topological defects that, by suppressing the magnetisation, enable the system to avoid a high kinetic energy cost.
The crossover between magnetised and non-magnetised states resembles a first-order phase transition, as both families of states are metastable close to the critical value of the spin-velocity coupling.
When heated up, systems prepared in a non-magnetised ground state display a maximum of the modulus of their magnetisation at a finite temperature.
This feature is reminiscent of an ``Order by Disorder''~\cite{Villain1980,Shender1996} (ObD) transition, in which a system develops a spontaneous magnetisation upon heating, and can be explained by the examination of the free energies of the various states of the system.
Finally, the magnetised states are characterised by a collective ordering of the instantaneous velocities, which is however rapidly destroyed by thermal fluctuations.
The above results are rather general. On one hand the choice of the spin-velocity coupling is highly constrained by the structure of the Hamiltonian dynamics. 
On the other hand neither the geometry nor the dimensionality seem to play a major role in our observations. 

The paper is organised as follows.
In Sec.~\ref{sec:Model} we introduce the model, emphasise the unusual features of its conservative dynamics, and describe the numerical simulations used thereafter.
In Sec.~\ref{sec:LowT} we characterise the low-temperature equilibrium states and predict their location in parameter space with simple energetic arguments, as well as their nature, resorting to a coarse-grained continuous theory.
In Sec.~\ref{sec:FiniteT} we consider the finite temperature regimes and describe an ObD phenomenon for a subset of values of the parameters.
The crossover towards order is studied, and proved to behave much like a first-order phase transition.
In Sec.~\ref{sec:Velocities} we discuss the alignment of velocities in collectively moving phases. 
Finally, we discuss the generality of our results, show that the choice of a spin-velocity coupling is highly constrained and present our conclusions regarding the link between this model and usual collective motion in Sec~\ref{sec:Conclusion}.

\section{The Model\label{sec:Model}}
The model, introduced in Ref.~[\onlinecite{Bore2016}], describes the dynamics of $N$ interacting particles that carry continuous planar spins of norm equal to $1$. 
The particles are confined to move in $2d$, in a periodic square box of linear size $L$. 
Their motion is described by the Lagrangian
\begin{align}
 \mathcal{L} &= \sum\limits_{i = 1}^{N} \frac{m}{2} \dot{\bm{r}}_i^2 + 
                \sum\limits_{i = 1}^{N}\frac{I}{2} {\dot{\bm{s}}_i}^2 +
                \sum\limits_{i = 1}^{N} K \dot{\bm{r}}_i \cdot \bm{s}_i
  - \frac{U_0}{2} \sum\limits_{k \neq i} U(r_{ik}) 
  + \frac{J_0}{2} \sum\limits_{k \neq i} J(r_{ik}) \cos\theta_{ik}
  \; ,
  \label{eq:Lagrangian}
\end{align}
that depends on position and spin variables.
The position of the $i$-th particle is denoted $\bm{r}_i$ and its velocity $\bm{v}_i = \dot{\bm{r}}_i$, while $r_{ik}= |{\bm r}_i - {\bm r}_k|$ is the distance between the centers of particles $i$ and $k$;$\theta_i$ is the angle that the spin forms with a reference axis and fully parametrises the continuous $2d$ spin $\bm{s}_i$ of unit modulus. 
The time derivative of the spin vector is indicated by $\dot{\bm{s}}_i$; 
$\theta_{ik}$ is the angle between the spin of particle $i$ and the one of particle $k$. 
Furthermore, $m$ is the mass of each particle and $I$ its moment of inertia.
$U$ is a short-ranged, isotropic and purely repulsive two-body interaction potential and $J$ is a short-ranged and isotropic ferromagnetic coupling between the spins. 
We characterise these two potentials below. 
For future convenience we made explicit the typical amplitudes
of the two-body potential, $U_0$, and the ferromagnetic coupling, $J_0$. 
$K$ is the parameter that controls the strength of the spin-velocity coupling, the term that yields the special properties to the model.
Throughout this paper, we will only consider the case $K\geq 0$ as the case $K \leq 0$ can be recovered by simply flipping all the spins.
In the following, we will introduce dimensionless variables according to the transformations $\bm{r}/\sqrt{I/m} \to \bm{r}$, $t/\sqrt{I/J_0} \to t$, $K/\sqrt{m J_0} \to K$, $L/J_0 \to L$, $U_0/J_0 \to U_0$ and, in the notation hereafter, we absorb $U_0$ in the definition of the two-body potential $U$.

In the Hamiltonian formalism the Lagrangian~(\ref{eq:Lagrangian}) transforms into the Hamiltonian
\begin{align}
   \mathcal{H} &= \sum\limits_{i = 1}^{N} \frac{1}{2} \bm{p}_i^2 + 
                \sum\limits_{i = 1}^{N}\frac{1}{2} \omega_i^2 -
                \sum\limits_{i = 1}^{N} K \bm{p}_i \cdot \bm{s}_i
  + \frac{1}{2} \sum\limits_{k \neq i} U(r_{ik}) - \frac{1}{2} \sum\limits_{k \neq i} J(r_{ik}) \cos\theta_{ik} \label{eq:Ham}
  \; ,
\end{align}
where we defined the canonical momenta $\omega_i = \dot{\theta_i}$ and $\bm{p}_i = \dot{\bm{r}} + K\bm{s}_i$.
As already stressed in Ref.~[\onlinecite{Bore2016}], this specific form of the momentum turns into a direct relation between the velocity of the centre of mass and the total magnetisation in the micro-canonical ensemble. 
In particular, setting the total momentum $\bm{P} = \sum_i \bm{p}_i $ to zero, the system spontaneously develops collective motion, whenever it magnetises. 
Its centre of mass velocity then satisfies $\bm{v}_G + K \bm{m} = 0,$  where
\begin{equation}
    \bm{m} = \frac{1}{N}\sum_{i = 1}^{N} \bm{s}_i \; ,
    \qquad\quad \mbox{and} \quad\qquad 
    \bm{v}_G = \frac{1}{N}\sum_{i = 1}^{N} \dot{\bm{r}}_i
    \; , 
\end{equation}
are the intensive magnetisation and the velocity of the centre of mass, respectively. 
Note that although free particles would tend to align their velocity on their own spin~\cite{Bore2016}, the collectively moving states at low energy that satisfy $\bm{P} = \bm{0}$ anti-align the particle velocities with their spins.
Also because the velocity of a magnetised state grows linearly with $K$, low potential energy moving states become very expensive in terms of kinetic energy as $K$ increases. 
As a result, for high enough values of $K$ non-magnetised states featuring topological defects were suggested as candidate ground states~\cite{Bore2016}. 
We study this hypothesis in detail in the body of the paper and will indeed validate it.\\

The soft interaction potentials are given by
\begin{align}
    J(r) &= (\sigma - r)^2 \Theta(\sigma - r) \; , \\
    U(r) &= U_0 (\sigma - r)^4 \Theta(\sigma - r) \; ,
\end{align}
where $\Theta$ is a Heaviside step function, $\sigma$ is the range of the interactions, that we
fixed to 1, and we took $U_0 = 4$ so that both potentials are equal at half range.\\

The numerical analysis is carried out using ``microcanonical'' Molecular Dynamics (MD) simulations
of the dynamics defined by the Hamiltonian in Eq.~(\ref{eq:Ham}), which conserves energy and momentum, and we restrict ourselves to initial conditions such that $\bm{P} = \bm{0}$.
We use random initial states with uniformly distributed $\left\{\bm{r}_i, \theta_i \right\}_{i=1, \dots, N}$ and $\left\{\bm{p}_i, \omega_i \right\}_{i=1, \dots, N}$ drawn from centered reduced Gaussian distributions. 
Such configurations are placed in a square box with periodic boundary conditions and, after giving some time for the dynamics to settle in, they are subjected to either a numerical annealing or a high-rate quench. 
We assume that the outputs of these microcanonical simulations can be interpreted from the point of view of canonical-ensemble statistical mechanics when sampling over initial conditions.
This equivalence supposes that the dynamics defined by the Hamiltonian~(\ref{eq:Ham}) are ergodic in phase space.
This ergodic hypothesis is often made in sufficiently interacting Hamiltonian systems, since the structure of their dynamics guarantees that Liouville's theorem applies, and that a flat distribution of microstates in phase space is stable when the dynamics are run.~\cite{Kardar2007}
While Liouville's theorem is not sufficient to guarantee ergodicity, here the hypothesis is supported by the results of Ref.~[\onlinecite{Bore2016}], where it is shown that mean-field canonical predictions for the magnetisation as a function of the energy and momentum qualitatively reproduce the numerical results obtained with MD simulations.
Further details on the simulation and numerical methods are provided in App.~\ref{app:Num}.
The phase diagram of this model has been fully characterised in the case $K=0$~[\onlinecite{Casiulis2019}]. 
Because aligned spins also attract each other, the spin fluid undergoes a liquid-gas phase separation at sufficiently low temperature and density~\cite{Casiulis2019}. 
In the present paper we concentrate on the physics of homogeneous phases and therefore set the density to $\rho = N/L^2 \approx 2.81$ (or, equivalently, $\phi \approx 0.55$ in terms of the packing fraction defined as $\phi = \rho \pi \sigma^2 / 16$).

\section{Finite-Size Ground States\label{sec:LowT}}

\subsection{The parameter space}

\begin{figure}
\centering
\includegraphics[width = .9\columnwidth]{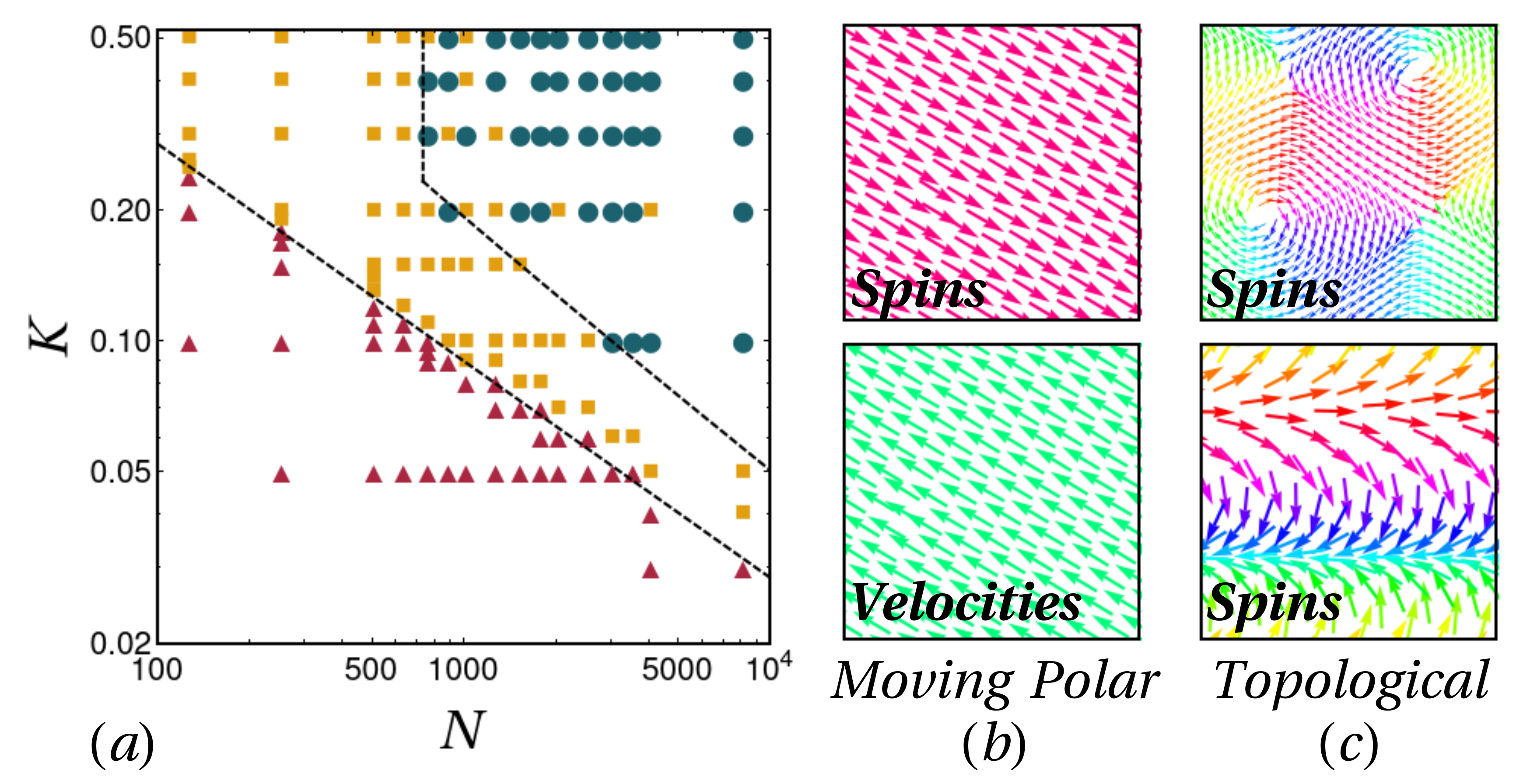}
\vspace{-2mm}
\caption{{\textbf{Zero-temperature homogeneous states.}}
    $(a)$ Nature of the ground states found numerically in the $K$-$N$ plane: polar states (red triangles), solitonic state (yellow squares),  vortex states (blue disks). Only the polar state is a collectively moving state. The nature of each point was determined using a majority rule over $1$ to $10$ different realizations of the annealing procedure. The crossover lines correspond to Eqs.~(\ref{eq:Ksol}), ~(\ref{eq:Nvor}) and ~(\ref{eq:Kvor}).
    $(b)$ Snapshots of a polar state obtained for $N = 128$, and $K = 0.2$, showing both the spins (top) and the corresponding velocities (bottom). The unit length for spin arrows is $K = 0.2$ times the one used for the velocities.
    $(c)$ Snapshot of a vortex state (top), obtained for $N = 768$, and $K = 0.4$, and of a solitonic state  (bottom), obtained for $N = 128$, and $K = 0.5$.
    }
    \label{fig:T0}
\end{figure}

In order to fully understand the conditions for the observation of moving states or topological defects, we start by focusing on equilibrium states in the lowest attainable energy configuration. They correspond to minima of the potential energy, which we interpret as states at very low temperatures and therefore call ``ground states''.
As shown in Fig.~\ref{fig:T0}, we find three kinds of ground states depending on the values of $K$ and $N$: \begin{enumerate}
    \item[(i)] {\it Polar states:} These are fully magnetised states with $m = 1$, with collective motion taking place in the direction opposite to the magnetisation as imposed by the momentum conservation, $\bm{v_G} = - K \bm{m}$.
    \item[(ii)] {\it Solitonic states:} These are non-moving states with $m \approx 0$, characterised by a continuous rotation of spins in one direction.
    \item[(iii)] {\it Vortex states:} These are non-moving states with $m \approx 0$, characterised by 4 topological point defects, 2 vortices and 2 anti-vortices.
\end{enumerate} 

Importantly enough, neither local nor global ordering of velocities is observed in the solitonic and vortex states. 
The reason for this is the high kinetic energy cost of the magnetised states, which increases  with $K$ and $N$ like $N K^2$. 
As these parameters are increased starting from a polar state, the system produces topological defects as a way of setting the global magnetisation to zero and therefore escape the collective motion imposed by momentum conservation. 
The strong frustration induced by the high kinetic energy cost associated to magnetised states with a finite velocity of the centre of mass has thus the effect of {\it generating} topological defects in the equilibrium ground states. 
Note that these defects are not thermal excitations and should not be confused with the vortices observed at the BKT transition in lattice models of XY spins.~\cite{Kosterlitz1973b,Tobochnik1979}
In fact we have shown in Ref.~[\onlinecite{Casiulis2019}] that no BKT transition occurs in the $K=0$ limit of the model~(\ref{eq:Ham}).

\subsection{Continuum theory \label{sec:Continuum}}
The nature of the observed ground states, can be further confirmed by considering the solutions of  a coarse-grained continuum theory for the magnetic properties of the model.
We define $\mathcal{L}_{kin}$,  $\mathcal{L}_{sv}$ and $\mathcal{L}_{int}$, the kinetic, spin-velocity and interaction contributions to the Lagrangian. 
They read
\begin{align}
    \mathcal{L}_{kin} &= \sum\limits_{i = 1}^{N} \left( \frac{1}{2} v_i ^2 + \frac{1}{2} \omega_i ^2 \right)
    \; , 
    \label{eq-app:kinetic}
    \\
    \mathcal{L}_{sv} &= K \sum\limits_{i = 1}^{N} \bm{v}_i \cdot \bm{s}_i 
    \; , 
    \label{eq-app:spin-velocity}
    \\
   \mathcal{L}_{int} &= - \frac{1}{2} \sum_{k \neq i} U(r_{ik}) + 
   \frac{1}{2} 
   \sum_{k \neq i} J(r_{ik})\bm{s_i \cdot s_k}
    \; , 
    \label{eq-app:interaction}
\end{align}
respectively, with $\bm{v}_i = \dot {\bm r}_i$ and $\omega_i = \dot \theta_i$. 
The symbol $\sum\limits_{k\neq i}$ indicates a double sum over $i$ and $k$ from 1 to $N$  with no repeated indices included.
Focusing first on the interaction term $\mathcal{L}_{int}$, we assume that the spins are locally well aligned, so that $\bm{s}_i \cdot \bm{s}_k \approx 1 - \theta_{ik}^2/2$, where $\theta_{ik} = \theta_i - \theta_k$, in the neighbourhood defined by the range of $J(r)$. Therefore, 
\begin{equation}
    \mathcal{L}_{int} \approx - \frac{1}{2} \sum_{k \neq i} U(r_{ik}) + \frac{1}{2}\sum\limits_{k\neq i}J(r_{ik}) - \frac{1}{4}\sum\limits_{k\neq i}J(r_{ik}) \theta_{ik}^2 \; , 
\end{equation}
The main effect of the first two terms is to ensure the uniformity of the density profile $\rho ({\bf r}) \approx \rho_0$. 
Accordingly, we will discard them in the following and concentrate on the last term, that we henceforth call $\mathcal{V}$, and want to approximate using a continuum description.
To do so, we assume that there exists a discrete mesh of allowed positions for the particles,~\cite{Dean1996} with arbitrarily small spacing, so that we may define coarse-grained fields on this mesh as: 
\begin{equation}
\mathcal{V} \equiv 
\sum\limits_{k\neq i} J(r_{ik}) \theta_{ik}^2 = \sum\limits_{\bm{r} 
} \sum\limits_{\bm{r'} 
} 
\sum\limits_{k \neq i} J(\bm{r},\bm{r'}) (\theta(\bm{r'}) - \theta(\bm{r}) )^2 \delta(\bm{r} - \bm{r_i}) \delta(\bm{r'} - \bm{r_k})
\; .
\end{equation}
with $\delta$ being Kronecker symbols. 
First, rewriting the sum over $\bm{r'}$ after the transformation $\bm{r'} \rightarrow \bm{r} + \bm{d}$, and using the rotational and translation invariances of $J$; second expanding the difference of $\theta$ to the first order, we obtain
\begin{equation}
\mathcal{V} \approx \sum\limits_{\bm{r}
} \sum\limits_{\bm{r+d} 
} 
\sum\limits_{k\neq i} J(d)
\, 
(\bm{\nabla}\theta (\bm{r}) \cdot \bm{d} )^2
\, 
\delta(\bm{r} - \bm{r_i}) \delta(\bm{r} + \bm{d} - \bm{r_k})
\; . \label{eq:V___cont}
\end{equation}
We now take the continuum limit by taking the mesh size to zero and find that in the absence of singular values of $\bm{\nabla} \theta$:
 \begin{equation}
\mathcal{V} \xrightarrow{a\rightarrow 0} \int d^2\bm{r} \, \rho(\bm{r}) \int d^2\bm{x} \,  \rho(\bm{r}+\bm{x})  J(x) (\bm{\nabla}\theta (\bm{r}) \cdot \bm{x} )^2 
\; . 
\end{equation}
For $J(r) = J_0 \Theta(\lambda - r)$, with $\Theta$ a step function, and recalling that $\rho ({\bf r}) \approx \rho_0$, we finally obtain, after integration of the second integral:
 \begin{equation}
\mathcal{V} \approx 
\int d^2\bm{r} \, \rho_0^2 \pi  J_0 \frac{\lambda^4}{4} (\bm{\nabla}\theta (\bm{r}))^2 \; . 
\end{equation}
The two other contributions to the Lagrangian, namely,  $\mathcal{L}_{kin}$ and $\mathcal{L}_{sv}$ are local, so that the same procedure as above can be trivially applied. Finally, we can write a continuum Lagrangian theory of the form
\begin{align}
    \mathcal{L}_c &= \int d^2\bm{r} \; 
    \rho_0 \left[ \frac{1}{2}\dot{\theta}^2(\bm{r}) -  \frac{c^2}{2}  (\bm{\nabla}\theta (\bm{r}))^2 + \frac{1}{2} v^2(\bm{r}) + K \bm{v}\cdot\hat{\bm{e}}(\theta (\bm{r}))  \right],
\end{align}
where $c^2 \equiv J_0 N_\lambda \lambda^2/16 $ with $N_\lambda \equiv \rho_0 \pi \lambda^2$ the number of particles in one interaction volume, and $\hat{\bm{e}}(\theta (\bm{r}))$ is a unit vector pointing in the direction given by $\theta$.
This form of the Lagrangian clearly shows that $K \bm{v}$ plays the role of a field, and creates a non-perturbative term even in the low-temperature expansion which otherwise gives a free theory for the XY model at low temperatures.~\cite{Kosterlitz1973b}

We can now write the Euler-Lagrange equations for $\theta$ and $\bm{v}$ to describe the low-energy regime of this theory, 
\begin{eqnarray}
   \bm{v} &=& - K \hat{\bm{e}}(\theta (\bm{r})) \label{eq:zeromom}
   \; , \\
   \left(\frac{\partial^2}{\partial t^2} - c^2 \triangle \right) \theta &=& \bm{v}\cdot\hat{\bm{e}}\left(\theta (\bm{r}) + \frac{\pi}{2}\right)
   \; ,
 \end{eqnarray}
 where the additional $\pi/2$ comes from the derivative of $\hat{\bm{e}}(\theta (\bm{r}))$ with respect to $\theta$.
Interestingly, at this level of description, Eq.~(\ref{eq:zeromom}) selects solutions such that the local momentum is zero.
When injecting this zero-momentum condition into the equation on $\theta$, we simply recover a wave equation on $\theta$: this is the usual spin-wave regime~\cite{Kosterlitz1973b} at low temperatures, that describes magnetised low-temperature states in finite size. 
To describe states that are close in energy to a finite-size polar moving state, we set $\theta(\bm{r}) = \theta_0 + \delta\theta(\bm{r})$, and $\bm{v} = - K \hat{\bm{e}}(\theta_0) + \bm{\delta v (\bm{r})}$ corresponding to a small perturbation around a polar state aligned along $\theta_0$, say $\theta_0 = 0$. The Lagrangian evaluated for such states reads,
\begin{align}
    \mathcal{L}_c 
    &= 
    \int d^2\bm{r} \; \rho_0 \left[ \frac{1}{2}\dot{\delta\theta}^2(\bm{r}) -  \frac{c^2}{2}  (\bm{\nabla}\delta\theta (\bm{r}))^2 + \frac{1}{2} v^2(\bm{r}) - K^2 \cos\delta\theta(\bm{r}) + K \bm{\delta v (\bm{r})}\cdot\hat{\bm{e}}(\delta\theta (\bm{r}))  \right]
    \;,
\end{align}
and contains a field term $K^2 \cos\delta\theta$ that is independent of the amplitude of the velocity perturbation.
In particular, it yields a term of order one, contrary to the other terms in the Lagrangian that are vanishingly small for $\delta v \sim \delta \theta \ll 1$.
Therefore, close to low-temperature polar ground states, the effective theory describing spin alignment is essentially a sine-Gordon theory with a field of amplitude $K^2$.

\begin{figure}
    \centering
    \includegraphics[width = .24\columnwidth]{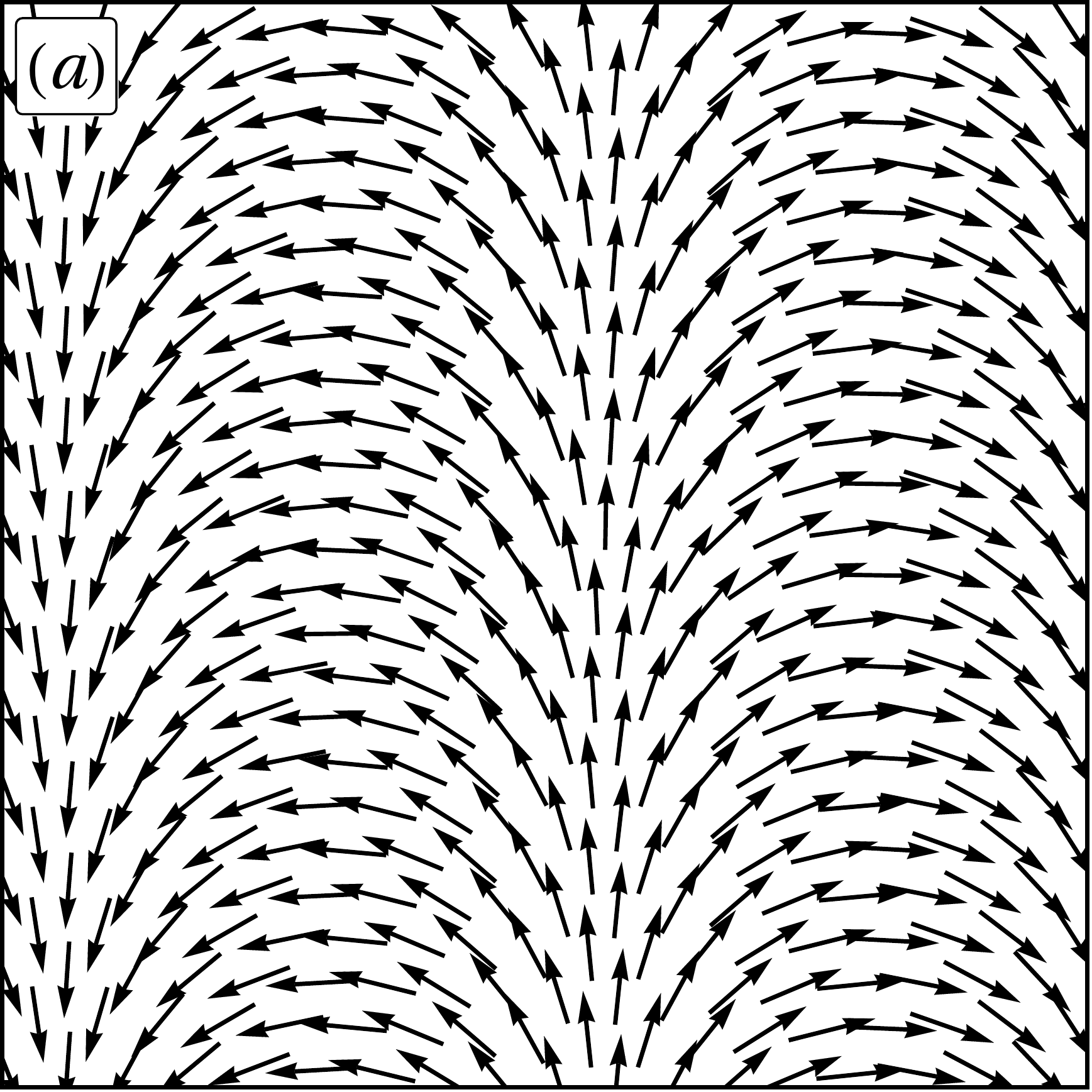}
    \includegraphics[width = .24\columnwidth]{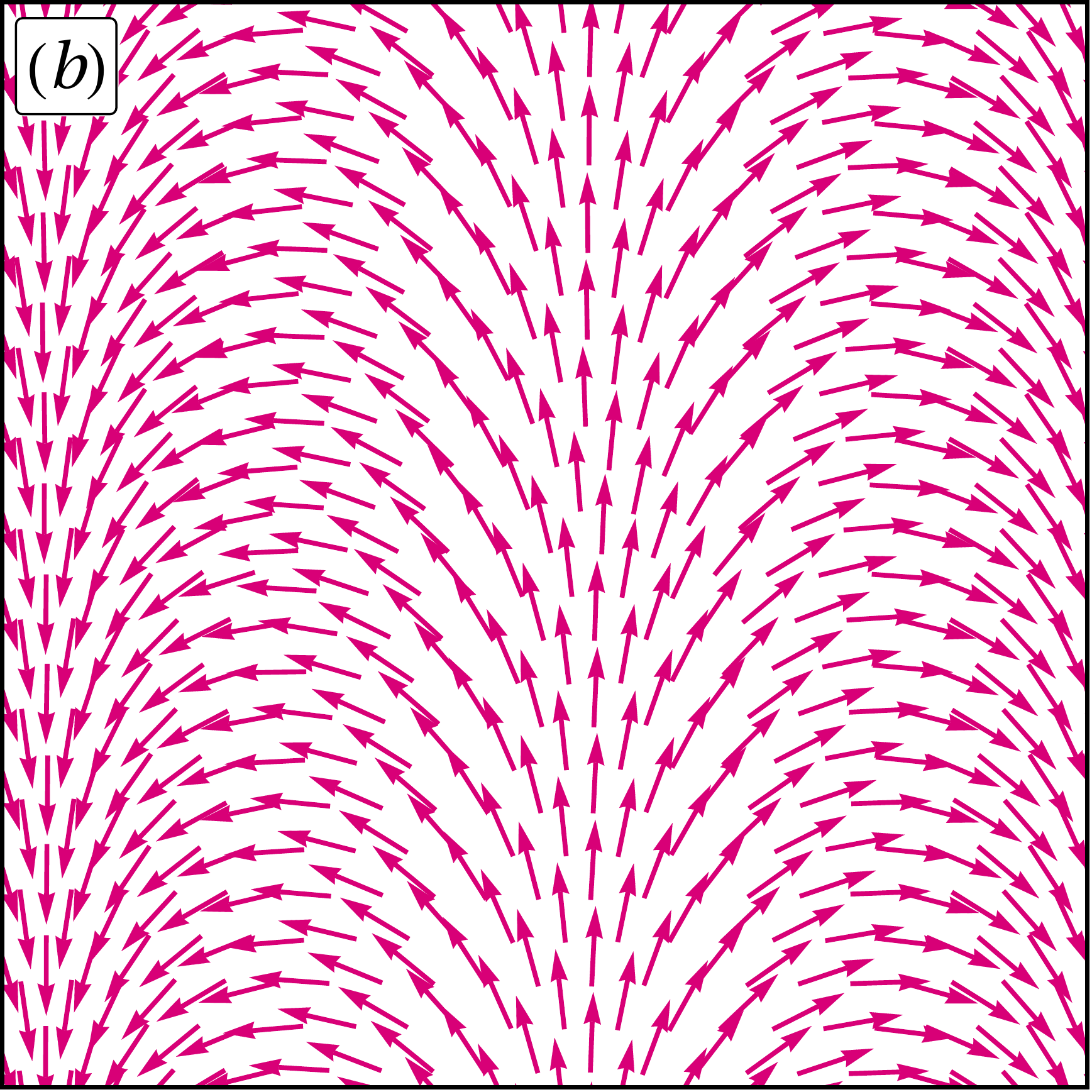}
    \includegraphics[width = .24\columnwidth]{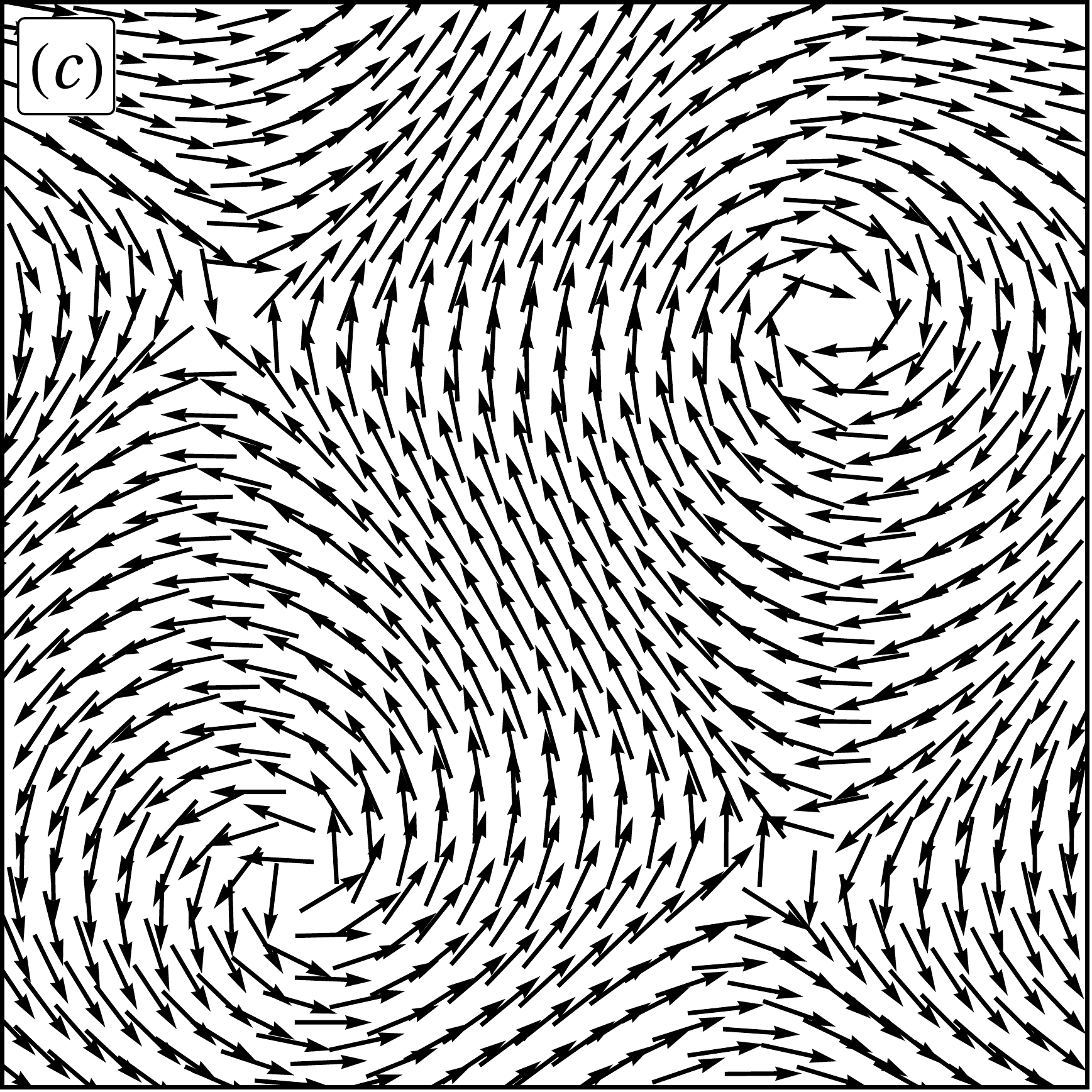}
    \includegraphics[width = .24\columnwidth]{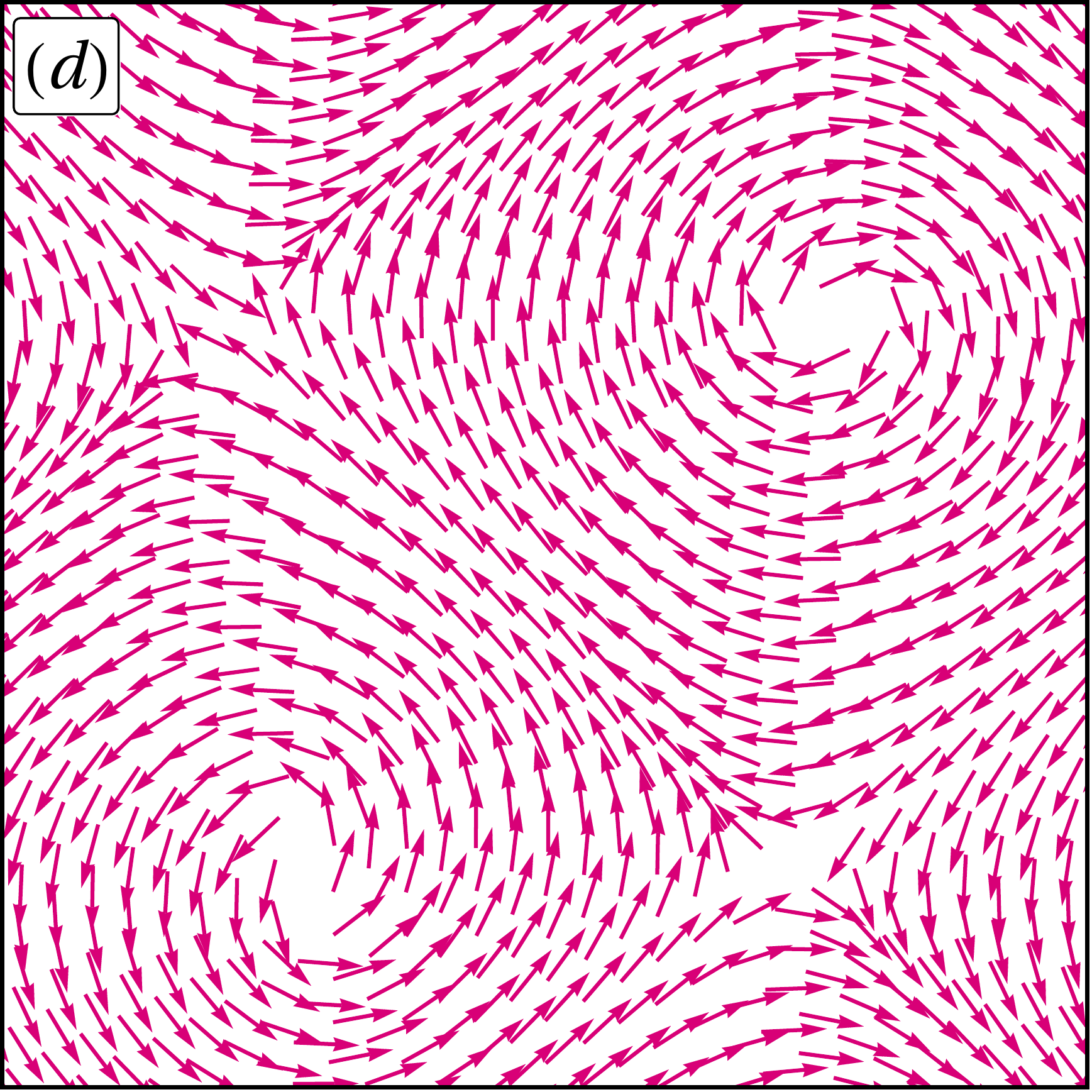}
    \caption{{\textbf{sine-Gordon solitons and vortices.}}
    $(a)$ sine-Gordon soliton found in the continuum theory, Eq.~(\ref{eq:Sol}), with $\Lambda = 1$, $x_0 = 0.5$, plotted in $[-1;1]^2$. 
    $(b)$ Ground state for $N = 512, K = 0.3$ found in numerical simulations.
    $(c)$ Laplace vortex-antivortex state in the continuum theory, Eq.~(\ref{eq:Laplace}), with $R = 0.295$, repeated over periodic boundary conditions and centered on the same points as in panel $(d)$. $(d)$ Ground state for $N = 768, K = 0.4$ found in numerical simulations.}
    \label{fig:sG}
\end{figure}

The sine-Gordon theory is known to feature solitonic solutions and vortex patterns, as reported in previous numerical works\cite{Barone71,Gouvea1990}.
We shall here briefly discuss both families of states.
Solitons are propagative solutions of the sine-Gordon theory whose functional form can be shown to be~\cite{Barone71}
\begin{align}
   \vartheta(\xi) &= 4 \chi \arctan\exp\left( \gamma \sqrt{h_{sG}}(\xi - \xi_0) \right) 
   - \chi \pi,
\end{align}
where $\vartheta$ is the orientation of the local magnetisation field, $\xi = x - c t$ is the usual variable of propagative solutions, with a propagation velocity $c$, $\xi_0$ is an offset for this variable that sets the position of the centre of the soliton, $\chi = \pm 1$ is the so-called polarity of the soliton, $\gamma = 1/\sqrt{1 - c^2}$ is a Lorentz factor associated to the propagation velocity, and $h_{sG}$ is the in-plane magnetic field of the sine-Gordon theory. 
This function represents a localised rotation of $2\pi$ of the magnetisation field along the $x$ direction.
Noticing that in simulations we only ever see one such rotation over the size of the whole box, we plot in Fig.~\ref{fig:sG}$(a)$ the magnetic field with a constant amplitude and an orientation defined by the following function of position $\bm{r} = (x,y)$,
\begin{align}
   \vartheta_\Lambda(x,y) &=\pi - 4 \arctan\exp\left( x - x_0\right) 
   - 4 \arctan\exp\left[ x - (x_0 - 2 \Lambda)\right]   
   - 4 \arctan\exp\left[ x - (x_0 + 2 \Lambda)\right] 
   \; ,
   \label{eq:Sol}
\end{align}
where $2\Lambda$ is the range of the box in which we plot the field, that is centered on $(0,0)$.
This field is a perfect reproduction of the arched states that we observe at low temperatures in the numerical simulations, as shown in Fig.~\ref{fig:sG}$(b)$, thus justifying the name ``solitons" that we give them.
As there is neither damping nor forcing in the present model, the soliton propagation velocity is a priori free to select any value.~\cite{Malomed1991} 
In practice, as demonstrated by the video shown in the Supp. Mat., we find that solitonic patterns do move in simulations, and that their propagation velocity fluctuates.

The sine-Gordon equation also features vortex solutions, but their exact functional shape is in general hard to derive. 
It is customary to consider vortex structures that are solutions of the Laplace equation, and to assume that they are little deformed by the addition of the non-linear term in the sine-Gordon equation~\cite{Shagalov92,Sinitsyn04}. 
One-defect solutions can then be written as
\begin{equation}
    \vartheta_{\pm}(x,y) = \pm \arctan\left(\frac{y - y_0}{x-x_0}\right) + \frac{R}{2} \log\left[ (y-y_0)^2 + (x - x_0)^2 \right], \label{eq:Laplace}
\end{equation}
where $(x_0,y_0)$ are the coordinates of the centre of the defect, $R$ is a parameter that tunes the twist of the spiral surrounding the defect, and the sign in front of the $\arctan$ determines whether the defect is a vortex ($+$) or an antivortex ($-$).
An example of a 4-vortex configuration obtained with this functional form is shown in Fig.~\ref{fig:sG}$(c)$, and compared to a 4-vortex numerical ground state shown in Fig.~\ref{fig:sG}$(d)$.
In this example, we chose the positions of the centers of the defects as well as the value of $R$ to best fit the numerical observations.

Altogether the low energy magnetic excitations of the continuous theory perfectly match the very low temperature magnetic states obtained numerically, confirming the intuition of the mechanism whereby the system escapes the kinetic energy cost of the moving polar state, when $K$ or $N$ are too large.
Note, however, that this description fails to capture the velocity configuration.
Indeed, in the continuum theory, we assume that the velocities (like the spins) are essentially all aligned.
While this perturbative approach enables us to predict topological magnetic excitations, that are quite different from a perfect polar state, it still predicts that the velocities should be opposed to the magnetisation field everywhere, whereas in simulations soliton and vortex patterns do not display any velocity alignment, as we further discuss in Sec.~\ref{sec:Velocities}.

\subsection{Energetic arguments}
One can further estimate the energy of these states and thereby obtain the range of parameters in which each state is observed.
The total kinetic and magnetic energy per particle in a moving polar state is 
\begin{equation}
E_{pol} = \frac{1}{2} K^2 - \frac{1}{2} z \bar{J} + E_U
\end{equation}
where the velocity is everywhere assumed to be $- K \bm{m}$, $z$ is the 
number of neighbours of each particle, and $\overline{J}$ is the mean value of the ferromagnetic coupling  for the inter-particle distances $\bar{r}$ in the considered configuration.
The energy contribution due to the repulsive interaction $E_U$ is likely to be very similar in the three different states of interest and therefore does not need to be evaluated.

The energy of a solitonic state can be approximated as follows. 
Let us consider that the rotation is smooth, so that two neighbours along the direction of the rotation (on average half the neighbours) are misaligned by an angle $2\pi \bar{r}/ L$, while spins are perfectly aligned along the other direction. 
Then, the total energy of a solitonic state can be approximated by
\begin{equation}
    E_{sol} \approx - \frac{z \bar{J}}{2} \left[\frac{1}{2}\cos\left(\frac{2\pi \bar{r}}{L}\right) + \frac{1}{2}\right] + E_U
    \; .
    \label{eq:energy-soliton}
\end{equation}
In both cases, we set the number of magnetic neighbours to 6 (in agreement with the numerical observations at the density used in the simulations), and compute the typical value of the ferromagnetic coupling by taking $\bar{J} = J(\bar{r})$. 
For a triangular lattice one has that $\bar{r} = \sigma / 2 \sqrt{\phi}$.

Equating now the two energies $E_{pol}$ and $E_{sol}$ yields an equation for the line separating polar and soliton ground states,
\begin{equation}
    K^{pol-sol}_c = \sqrt{\frac{z \bar{J}}{2}  \left( 1 -  \cos \frac{2\pi \bar{r}}{L_c}\right)}
    = \sqrt{\frac{z \bar{J}}{2}  \left( 1 -  \cos \left[2\pi \bar{r} \sqrt{\frac{ \rho}{N_c}}\right]\right)}
    \; , \label{eq:Ksol}
\end{equation}
which is the lower dashed black line reported in Fig.~\ref{fig:T0}. 
It captures remarkably well the crossover between the two kinds of states.
Note that approximating the cosine by its Taylor expansion up to second order, yields $K^{pol-sol}_c L_c \approx \sqrt{\pi^2 z \bar{J} \bar{r}^2}$, and stresses that $K$ and $L$ play a similar role in breaking the magnetic order.

To predict the curve that separates the solitonic and vortex states, we use an exact expressions for the excess energy of a vortex-antivortex pair compared to a fully magnetised configuration in an on-lattice XY model~\cite{Kosterlitz1973b}. 
Taking into account only nearest-neighbour pairs of defects, this excess energy can be written as 
\begin{equation}
\Delta E^{2dXY}_{4D-pol} = E^{2dXY}_{4D} - E^{2dXY}_{pol} \approx 8 \pi \bar{J} \ln\frac{d}{a} 
\; , 
\end{equation}
where $d$ is the distance between the two defects and $a$ is the size of the centre of the defect. 
Here $d=L/2$ and the vortex core diameter is of the order of the particle-particle spacing $a\simeq \bar{r}$. 
We then equate $\Delta E^{2dXY}_{4D-pol}$ to $E_{sol} - E_{pol}$ in its Taylor-expanded form and find a critical value for the number of particles above which four vortices states are more favorable than the solitonic ones: 
\begin{align}
    N^{sol-vor}_c &= \exp\left( 2 \phi z \right) 
    \; . 
    \label{eq:Nvor}
\end{align}
This corresponds to the vertical line in Fig.~\ref{fig:T0}. 
While it does correspond to the size above which we mostly observe vortices, we note that $K$ also plays a role in the change between soliton and vortex states that is not explained by this crude estimate of the energy. 
A way to take into account the effect of $K$ on the vortex state energy is to reckon that $K$ plays the same role as an in-plane field in the magnetic model analog, and to use the expression for the energy of a vortex-antivortex pair in a field:~\cite{Gouvea1990} 
\begin{equation}
    \Delta E^{h}_{4D-pol} \approx E_{4D}^{2dXY}(h) - E_{pol}(h) = 
    \Delta E^{2dXY}_{4D-pol} + 8\pi^2 \sqrt{h}\left(\sqrt{\frac{d}{a}} - 1 \right)
    \; . \label{eq:Vortexcost}
\end{equation}
In the magnetic analogue $h$ is the amplitude of an external in-plane field, while here it is an effective field due to the spin-velocity coupling $K$.
According to the effective sine-Gordon theory (see Sec.~\ref{sec:Continuum}), $h = K^2$.
An alternative and more intuitive manner to obtain the quadratic dependence on $K$ is to recall that Eq.~(\ref{eq:Vortexcost}) gives the excess energy of vortex-antivortex pairs with respect to a magnetised configuration under the same field. 
As a consequence, both energies should in principle be written in the same frame. 
In our case, however, we are comparing the energy of a non moving vortex configuration to that of a moving polar phase. 
We should therefore translate all velocities by $\bm{v_G} = - K \bm{m}$. 
This change of reference frame creates, however, a new $K$-dependent term in the Lagrangian $\mathcal{L}_{Gal}  = - K^2 \bm{m} \cdot \sum_{i = 1}^{N}\bm{s}_i$, which contains an effective field on spins of magnitude $K^2$.
Finally, replacing the field in the expression of the in-field excess energy of vortices and equating $E_{4D}^{2dXY}$ to $E_{sol}$ yields the following equation with a $K$ dependence:
\begin{equation}
    K^{4D-pol}_c = \frac{2 \phi z - \ln\sqrt{N_c}}{\pi \left(\sqrt{N_c} - 1\right)}. \label{eq:Kvor}
\end{equation}
The resulting line is shown, when it is higher than the zero-field one, in Fig.~\ref{fig:T0}.
It fits numerical observations well, even though the crossover between solitons and vortices is rather broad as we are working with finite and moderately small system sizes.
Note that Eqs.~(\ref{eq:Ksol}) and~(\ref{eq:Kvor}) predict that the soliton states disappear at a finite value of $N$: this is expected as solitons carry a non-zero topological charge (the vector field winds once around one direction of the Euclidean torus), and should not survive in the thermodynamic limit, unlike pairs of vortices and antivortices (the total winding number of which is $0$).
Furthermore, the number of vortices or solitons, is \textit{a priori} independent of the size of the system.
Indeed, while defects usually appear in equilibrium due to local frustration, such structures here develop only because the \textit{total} magnetisation should be zero for the kinetic energy to be low, resulting in a global frustration between the conservation of the momentum and the one of the energy.
As a result, there is no typical length scale associated to frustration smaller than the size of the system itself, and the $m = 0$ patterns are scale-free.
One can also reason from the point of view of energetic costs: additional vortices or solitons would only create more local curvature, leading to an increased magnetic cost, but no kinetic energy gain.
Finally, one could \textit{a priori} circumvent defect formation by scaling $K$ with $N$, in the same way that one usually rescales interaction constants in mean-field couplings.
Here, the right choice would be $K \sim N^{-1/2}$ so that the kinetic energy of a magnetised state remains unchanged when increasing $N$.
However, we choose not to discuss such scalings at length in this paper, for two different reasons.
First, since the collective speed in this system scales like $K$, such a coupling would still lead to vanishingly small velocities.
Second, such interactions, while they are used in fully-connected descriptions, are rather artificial, and would take this model even further from realistic microscopic models.

\section{Finite Temperature States\label{sec:FiniteT}}
\subsection{magnetisation as a function of Temperature\label{sec:MagProp}}
Since we are performing microcanonical simulations, statistics can only be obtained at one value of the energy by sampling initial conditions.
Rather than using the energy as a control parameter, we assume ergodicity and treat the system as one of statistical mechanics, as explained in Sec.~\ref{sec:Model}.
In particular, we introduce the statistical temperature $T$, which is usually measured using velocity fluctuations.
In the present case the coupling between the velocities and the spins modifies, however, the usual equipartition relations:~\cite{Bore2016}
\begin{eqnarray}
T &=& \frac{1}{N} \sum_{i=1}^N \omega_i^2 - \left( \frac{1}{N} \sum_{i=1}^N \omega_i\right)^2
\label{eq:rotation-temp}
\; , 
\\
2 T &=& \frac{1}{N} \sum_{i=1}^N \bm{p}_i^2 - \left( \frac{1}{N} \sum_{i=1}^N \bm{p}_i\right)^2 - K^2\left(  \frac{1}{N} \sum_{i=1}^N \bm{s}_i^2 - \left( \frac{1}{N} \sum_{i=1}^N \bm{s}_i\right)^2\right) \label{eq:EquipMom}
\;.
\end{eqnarray}
The latter expression shows that the fluctuations of momenta and spins are coupled as long as $K \neq 0$, in such a manner that the temperature is not directly proportional to the translational kinetic energy. 
The former, on the other hand, shows that the temperature is proportional to the rotational kinetic energy, which is thus the most natural thermometer.
Having checked that both definitions yielded the same mean value of the temperature, we heretofore only refer to the mean temperature over independent configurations at the same energy, $\langle T \rangle$, as $T$.\\

\begin{figure}
    \centering
    \includegraphics[width = .4\columnwidth]{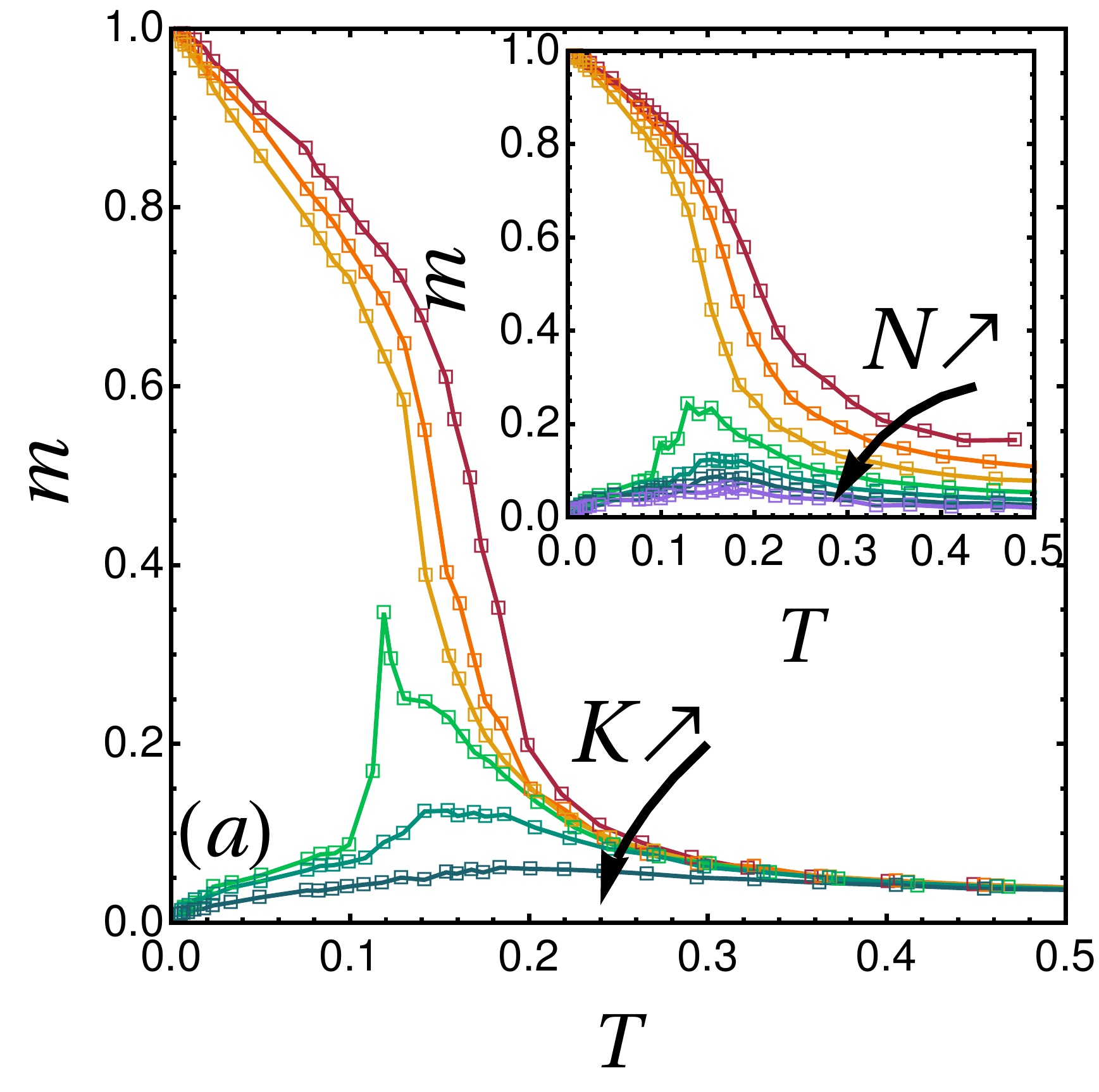}
    \includegraphics[width = .415\columnwidth]{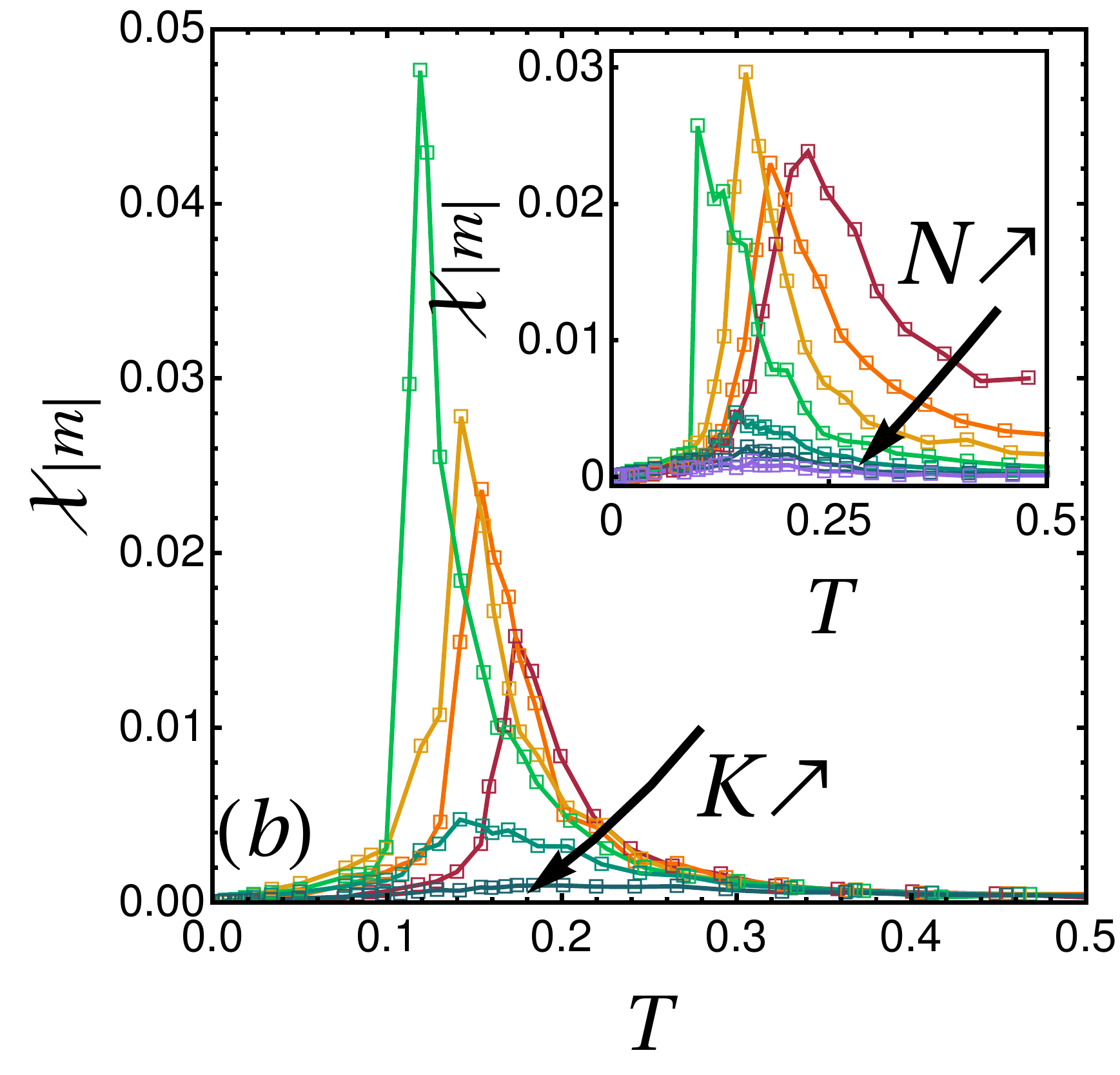}
    \caption{{\textbf{Finite-temperature behaviour and Order by Disorder.}}
    $(a)$ Magnetisation modulus versus temperature for $K = 0, 0.05, 0.06, 0.07, 0.1, 0.2$ at $N = 2048$ (main plot), and for $N= 128, 256, 512, 1024, 2048, 4096, 8192$ at $K = 0.1$ (inset). $(b)$ The reduced susceptibility per particle versus temperature for the same parameters as in panel $(a)$. 
    }
    \label{fig:T}
\end{figure}

We start by studying the effects of the finite temperature on the magnetic properties of the homogeneous phases of the system using the (intensive) magnetisation modulus $ m = \left\langle \left| {\bf m} \right| \right\rangle$, which is the order parameter of the polar phase.
Fig.~\ref{fig:T}$(a)$, displays the equilibrium value of $m$ against $T$ for increasing values of $K$ at fixed $N$ (main panel) or increasing values of $N$ at fixed $K$ (inset).
When varying $K$, we observe three types of behaviours.

\begin{itemize}
\item[(i)]  For small $K$, when the ground state is the moving polar state, we observe a finite-size crossover between a (moving) ferromagnetic state at low temperatures and a non moving paramagnetic state at high temperatures (red-to-orange curves of Fig.~\ref{fig:T}$(a)$). This is the same crossover as the one reported in the case $K=0$ between the magnetised ground state and the paramagnetic high temperature phase: the polar state does not survive in the thermodynamic limit in $2d$ due to low energy spin waves excitations.~\cite{Casiulis2019}
The effect of $K$ is to decrease the value of the magnetisation modulus at finite temperatures with respect to the $K=0$ case.
This can be understood by analyzing the spin waves excitations at low temperature within the framework of the continuum-limit theory described in Sec.~\ref{sec:Continuum}.
Indeed, close to $T = 0$ these spin-waves are described by a free Gaussian theory, that comes from the Taylor expansion of the spin-spin interaction terms in the Hamiltonian at second order.~\cite{Kosterlitz1974}
Spin waves suppress global magnetisation as the temperature increases. Furthermore, the effective spin-spin coupling decreases with temperature when taking into account higher-order terms in the Taylor expansion.
In the $K\neq 0$ case, an effective field is generated on top of the spin-spin interaction term. The direction of this effective field is opposite to the one of global magnetisation, and its amplitude grows with $K$.
As a result, the effective spin-spin coupling is further decreased by the addition of $K$, leading to stronger spin-wave suppression of the magnetisation as $K$ increases.
In short $K$ favors the bending of the magnetisation field.
\item[(ii)]
For large $K$, when the ground state has zero magnetisation modulus because of its topological structure, the magnetisation modulus crosses over from $m\simeq 0$, to a high-temperature paramagnetic regime where $m \sim 1/\sqrt{N}$.
\item[(iii)]
When $K$ takes moderately large values, although the ground state remains non polar ($m(T=0) = 0$), the magnetisation grows to a maximum at a finite temperature, before it crosses over to the paramagnetic regime at high temperatures. We shall focus on this intriguing phenomena in the next section~\ref{sec:ObD}.
\end{itemize}
When increasing $N$ at a fixed $K$, as shown in the inset of Fig.~\ref{fig:T}$(a)$, the same three scenarii are observed.
This is another sign of the qualitative similarity between the roles played by $K$ and $N$, that was already pointed out in Sec.~\ref{sec:LowT}.\\

It is also customary in the study of XY spins to introduce a ``reduced" susceptibility associated to $m$,~\cite{Tobochnik1979,Archambault1997,Casiulis2019}
\begin{equation}
    X_{|m|} = N \left(\langle m^2 \rangle - \langle |m| \rangle^2 \right).
\end{equation}
While it is not the usual response function of a magnetic system to an external field, this quantity captures the broadening of the distribution of $|m|$ that occurs at the finite-size paramagnetic-ferromagnetic crossover with a peak that grows and narrows with the system size.~\cite{Archambault1997}
In Fig.~\ref{fig:T}$(b)$, we plot the ``reduced'' susceptibility per particle, $\chi_{|m|} =  X_{|m|}/N$ against the temperature $T$ defined in Eq.~(\ref{eq:rotation-temp}), corresponding to the magnetisation curves shown in Fig.~\ref{fig:T}$(a)$.
Upon increasing $K$ at a fixed $N$ (main panel), the susceptibility peak observed in the case of a polar ground state grows with $K$ and is shifted towards lower temperatures. The same behaviour is observed upon increasing $N$ at $K=0$~\cite{Casiulis2019}.
A peak is also observed in cases in which the magnetisation features a local maximum [green curve in Fig.~\ref{fig:T}$(b)$], and it is even higher and shifted to lower temperatures compared to the peak observed for polar states.
Finally, the peak is suppressed at $K$ large enough, when the magnetisation simply crosses over from the one of a topological ground state to that of a paramagnet.
Likewise, when varying $N$ at a fixed value of $K$ (inset), the same scenario is observed: the peak of the extensive susceptibility $X_{|m|} = N \chi_{|m|}$ grows, is shifted to lower temperatures, and is then suppressed.
This behaviour is in sharp contrast with the smooth, algebraic evolution of the height of the peak reported in the case $K = 0$.~\cite{Casiulis2019}

\subsection{Unusual Order-by-Disorder Scenario\label{sec:ObD}}

As described in Sec.~\ref{sec:MagProp}, a surprising feature of the model is that, for some finite value of $K$ or $N$, the magnetisation exhibits a local maximum at a finite temperature.
In other words, the  order parameter {\textit{grows}} when thermal fluctuations are switched on. This behaviour is akin to Order-by-Disorder transitions~\cite{Villain1980,Shender1996,Guruciaga2016}
observed in frustrated magnets, whereby a system with a non-trivially degenerate ground states develops long-range order by the effect of classical or quantum fluctuations. 
In such geometrically frustrated spin models, the ObD phenomenon is due to a huge disproportion in the density of low-energy excitations associated with a specific ground state. Here the mechanism must be of a different nature.

\begin{figure}
    \centering
    \includegraphics[width = .3\columnwidth]{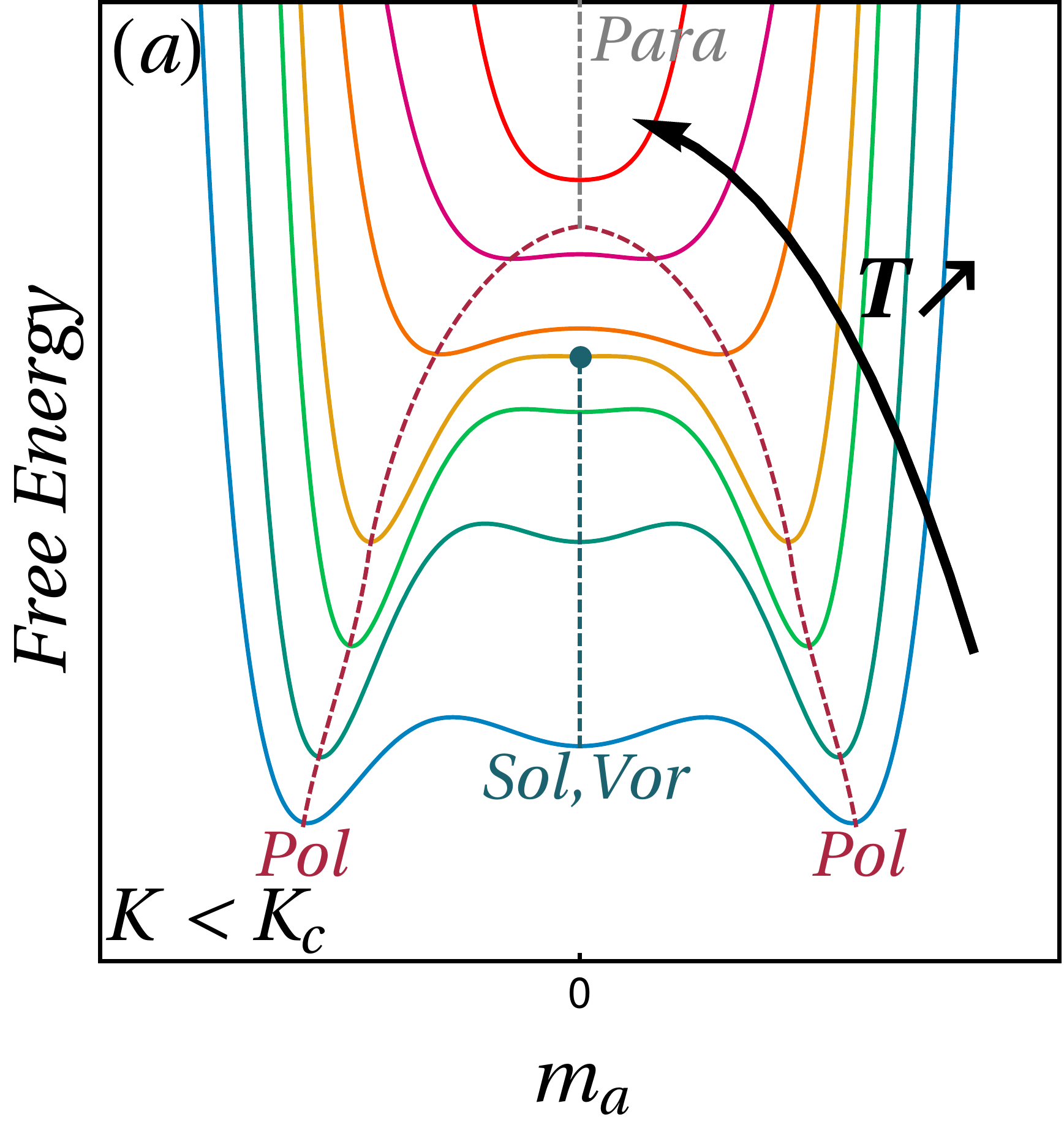}
    \includegraphics[width = .3\columnwidth]{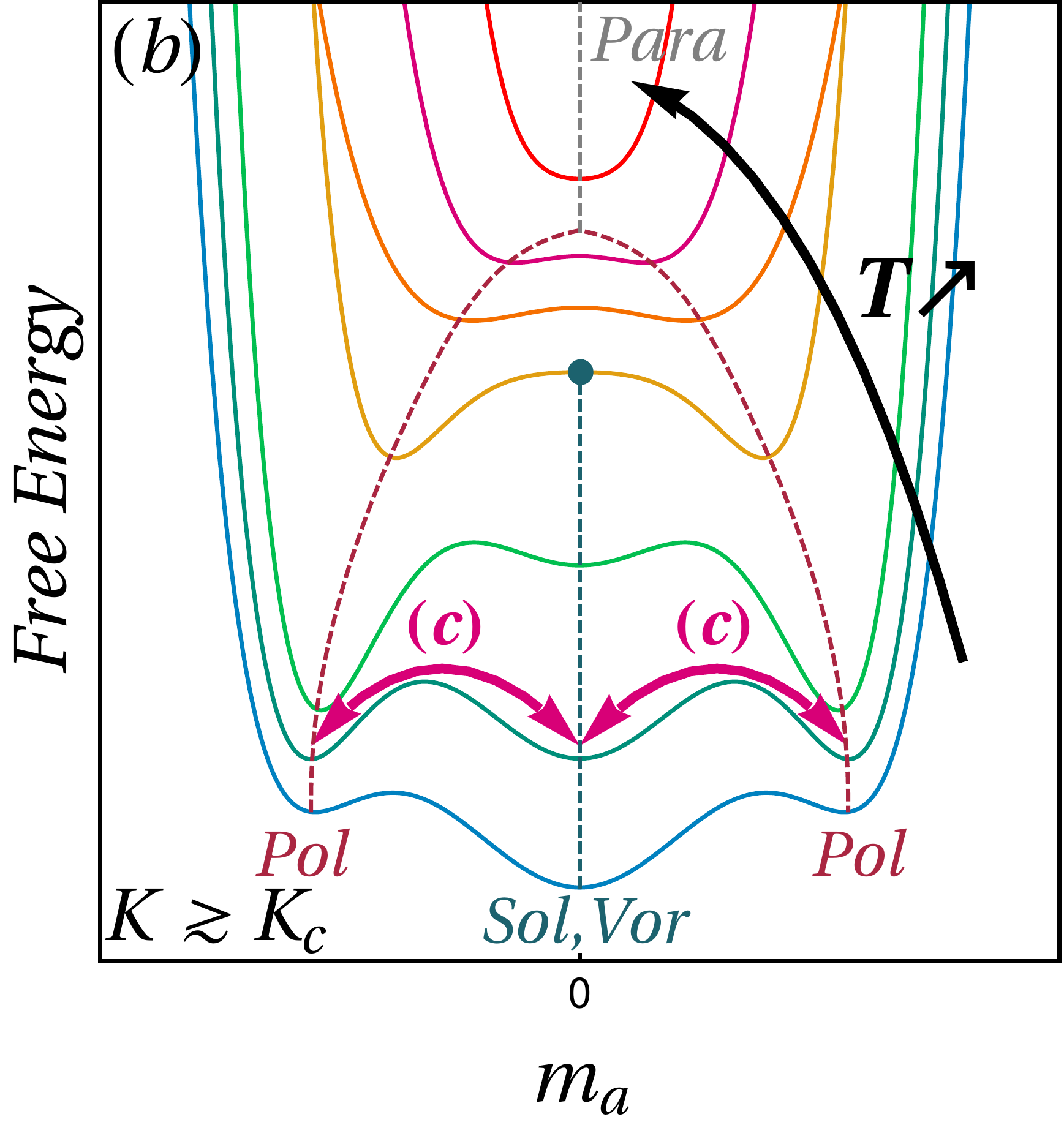}
    \includegraphics[width = .32\columnwidth]{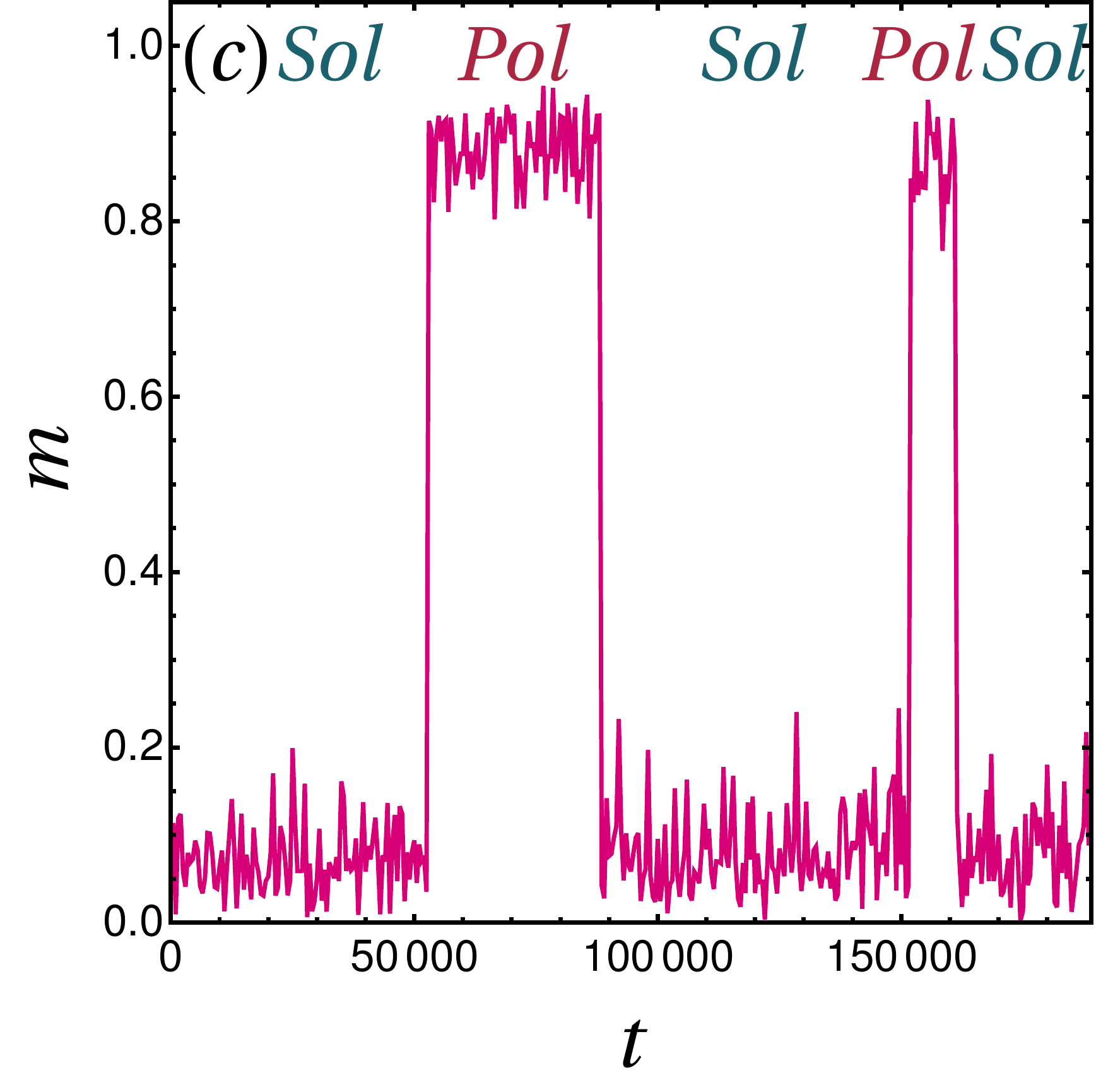}\\
    \includegraphics[width = .96\columnwidth]{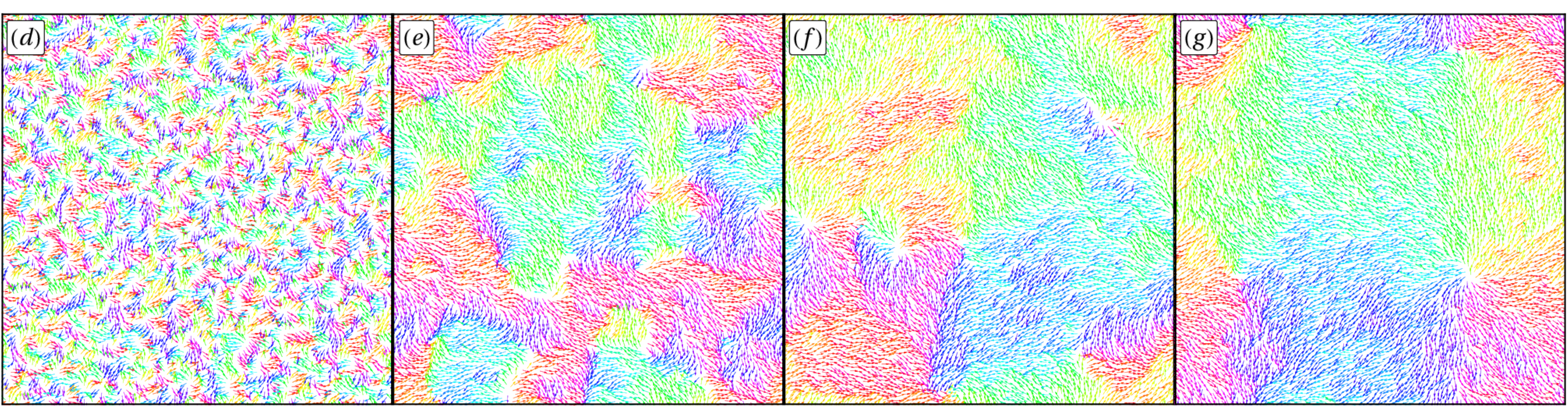}
    \caption{{\textbf{Coexistence and metastability.}}
    $(a)$ Sketch of the evolution of the free energy as a function of the magnetisation (along a particular direction) with increasing temperature (bottom to top) for low values of $K$: the low-temperature global minima are polar ($Pol$) and simply follow a Mexican hat scenario until they merge into a paramagnetic ($Para$) state, while soliton and vortex ($Sol,Vor$) states are only metastable at low temperatures.
    $(b)$ Free energy for $K \geq K_c$: the topological states are now global minima of the free energy until they cross the locally stable polar states at a finite temperature.
    At an intermediate temperature, when the $Pol$ and $Sol,Vor$ states have about the same free energy, switches between the two minima are observed in the Hamiltonian dynamics, as shown in $(c)$.
    $(c)$ Switches between low- and high- magnetisation regimes in the course of time for $N = 128, K = 0.25, T = 0.03$.
    $(d)-(g)$ snapshots after a quench of $N = 8192$ particles at $K = 0.03$ from $T \approx 2$ to $T \approx 0.04$, at times $t = 50, 150, 250, 4000$ after the quench, respectively. 
    The different colours represent different local magnetic orientations.
    }
    \label{fig:Crossover}
\end{figure}

In fact, in the present case the growth of a spontaneous magnetisation upon increasing the temperature is essentially due to the fact that the kinetic energy cost associated to collective motion in polar states decreases with $T$.
The reason is that, as already discussed in Sec.~\ref{sec:LowT}, the magnetisation modulus decreases as the temperature is increased due to spin waves fluctuations, and so does the kinetic energy cost induced by momentum conservation.
Hence, at moderately large $K$ magnetised states become less and less costly compared to the soliton or vortex ones upon increasing the temperature, until a point where the free energy of the (moving) polar states crosses the one of the (non moving) soliton or vortex states and become the preferred minimum, in a way similar to a first-order phase transition. 
When the temperature is further increased, the topological states become unstable, and the polar minima continuously merge into a single paramagnetic minimum, as in the $K = 0$ case.~\cite{Casiulis2019}
This scenario is pictorially sketched in Fig.~\ref{fig:Crossover}$(a)-(b)$ in terms of a Landau-like free energy representation. 
The fact that the system develops a spontaneous magnetisation and exhibits collective motion with a finite velocity of the centre of mass, upon increasing the temperature starting from a topological ground state with zero magnetisation, is thus due to the re-entrant shape of the surface separating the polar states from the solitonic ones in the extended phase diagram of Fig.~\ref{fig:T0}, when a vertical axis corresponding to temperature is added.

Note that this interpretation relies on the existence of metastable states, corresponding to local stable minima of the free energy at finite temperature. In the thermodynamic limit, metastablity is forbidden in finite dimensional systems due to the convexity of the free energy. 
The ObD-like transition described here is in fact only a smooth crossover, and is continuously suppressed as the system size is increased. 
Yet, at finite $N$ one can explicitly check in numerical simulations that both the topological and polar states are indeed locally stable at low enough temperature, as explained below.
First, we fine-tune $K$ and $N$ to get a system very close to the polar-soliton ground-state crossover, and we choose $N$ to be rather small (here $N = 128, K = 0.25$).
We then let the dynamics run at a finite but low temperature, $T = 0.03$.
As shown in Fig.~\ref{fig:Crossover}$(c)$, switches between low- and high-magnetisation states can be observed, proving that both states are indeed local minima of the free energy. 
However, even for such small systems, one must wait very long times for switches to happen: this is an indication that the energy barrier between the two states is rather high compared to thermal fluctuations.
One indeed expects that the energy barrier between a solitonic pattern and a polar state should scale linearly with the system size, since a collective reorganization of the spin degrees of freedom is necessary to switch between them.

Another indication of the existence of topological non-magnetised states beyond the range of parameters for which they are the equilibrium configurations is provided by the following observation.
Consider a system with $K$ and $N$ such that at zero temperature the ground state is the polar one, and quench it from a high-temperature to a very low temperature.
The result, shown in Fig.~\ref{fig:Crossover}$(d)-(g)$, shows that because non-equilibrium vortices appear during such quenches (as already observed for $K = 0$~\cite{Casiulis2019}), the system often ends up in a 4-vortex state (whose centers being located at the equidistant points where red, yellow, green and blue regions meet).

\subsection{Qualitative Free Energy Expression}
\begin{figure}
    \centering
    \includegraphics[width= .4\columnwidth]{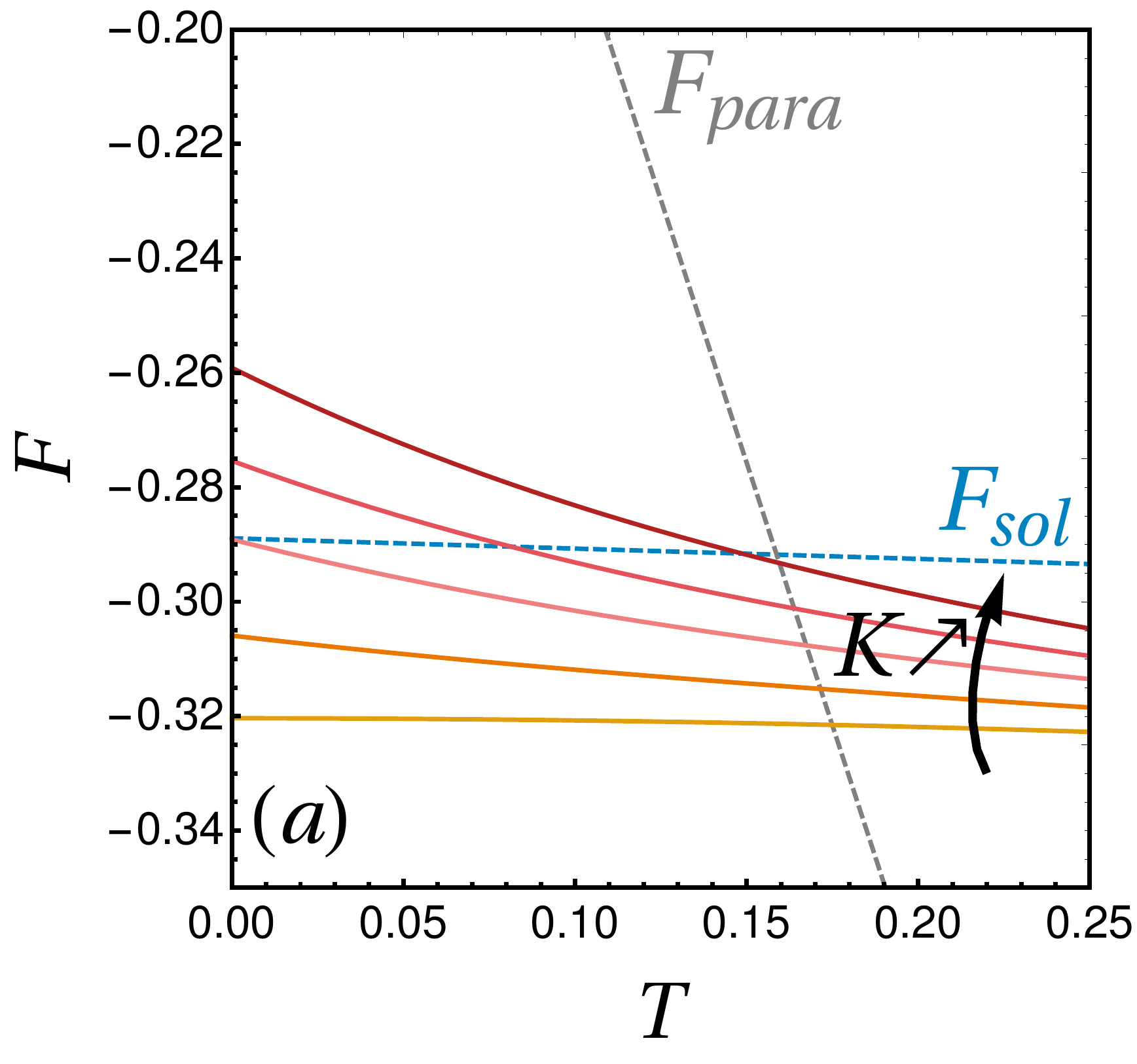}
    \caption{\textbf{Low-temperature free energies of magnetic and solitonic branches.}
    Free energies $F_{sol}$ (dashed blue line) and $F_{para}$ (dashed gray line) and $F_{pol}$ (full lines with different values of $K$ between $0$ (yellow) and $0.35$ (red)) defined in the text versus temperature, for $N=128$.
    }
    \label{fig:FreeEnergies}
\end{figure}
To go beyond the sketchy description of the free energy used in Fig.~\ref{fig:Crossover}, we introduce simple expressions for the free energy densities of both families of states at low temperatures and in spatially homogeneous states, in the same spirit as the discussion of the energy of the different ground states (Sec.~\ref{sec:LowT}).
First, the energy per particle of a soliton will be considered to be unaltered by temperature, and to remain of the same form $E_{sol}$ as in Eq.~(\ref{eq:energy-soliton}).
Then, we approximate the associated entropy per particle, $S_{sol}$,  by considering that it is only associated to the number of ways in which one can draw a soliton pattern of size $L$ with $N$ particles, separated from their nearest neighbours by a typical distance $\bar{r}$,
\begin{equation}
    S_{sol} = \frac{1}{N}\ln\left(\frac{L}{\bar{r}}\right)
    \; .
\end{equation}
The energy of a polar moving state at finite temperature should depend on the total magnetisation of the system to reflect the momentum conservation constraint.
Therefore, we modify the expression of $E_{pol}$ introduced in Sec.~\ref{sec:LowT} to include this effect,
\begin{align}
    E_{pol} &= - \frac{1}{2}z\bar{J} + \frac{1}{2}K^2 m(K,N; T)^2
    \; ,
\end{align}
where $m(K,N; T)$ is the intensive magnetisation modulus.
The exact form of $m$ at all temperatures and for all values of $K$ below $K_c$ is not known exactly. 
We thus choose to drop the $K$ dependence and to approximate $m$ by the exact expression obtained within a spin-wave approximation~\cite{Tobochnik1979,Archambault1997},
\begin{align}
    m(N; T) &= \exp\left( - \frac{T}{8 \pi \bar{J}} \ln(b N) \right)
    \; ,
\end{align}
where $b$ is a constant that was evaluated from finite-size scaling~\cite{Casiulis2019} in the case $K = 0$, yielding the value $b\approx 5.5$.
Finally, assuming that the system is well described by the spin-wave approximation in the considered domain of temperatures, we approximate the entropy per particle of magnetised states by counting how many ways there are to fit waves with a typical wave-length in an $L\times L$ box,
\begin{align}
    S_{pol} &= \frac{1}{N}\ln\left( \frac{L}{\xi_{sw}(N;T)}\right)^2
    \; ,
\end{align}
where we took the typical wave-length to be the spin-wave correlation length $\xi_{sw}$, that itself is well approximated, at low temperatures, by~\cite{Tobochnik1979}
\begin{align}
    \xi_{sw} &= L m(N; T) 
    \; .
\end{align}
We then have simple approximations for the free energies per particle $F_{pol}$ and $F_{sol}$ associated to polar and soliton states at low temperatures:
\begin{eqnarray}
    F_{pol} &=& - \frac{1}{2}z\bar{J} + \frac{1}{2}K^2 m(K,N; T)^2 +  \frac{2 T}{N}
               \ln m(N;T)
               \; ,
               \label{eq-app:free-mag}
               \\
    F_{sol} &=& - \frac{1}{2}z \bar{J} \left[\frac{1}{2}\cos\left(\frac{2\pi \bar{r}}{L}\right) + \frac{1}{2}\right] - \frac{T}{N}\ln\left(\frac{L}{\bar{r}}\right)
    \; .
    \label{eq-app:free-sol}
\end{eqnarray}
We can also approximate the free energy per particle of paramagnetic states by assuming that their magnetic energy averages to $0$, and that their entropy is simply given by the choice of random angles for all particles, so that
\begin{align}
    F_{para} &= - T \ln 2\pi
    \; . 
    \label{eq-app:free-para}
\end{align}
The expressions (\ref{eq-app:free-mag}), (\ref{eq-app:free-sol}) and (\ref{eq-app:free-para}) for the different branches of interest of the free energy density can be used to estimate which state is favoured at a given temperature for a given value of $K$ and $N$.
In particular, we can determine parameters such that $F_{sol} = F_{pol}$ for $K \geq K_c$.
Indeed, at the temperature such that these two branches cross, the system goes from a low-magnetisation soliton to a magnetised state. 
The crossing between $F_{sol}$ and $F_{pol}$ therefore corresponds to the temperature $T_{max}$ at which a maximum of the magnetisation is reached, with height $m_{max}$.
The results obtained when varying $K$ for $N = 128$, as in Fig.~\ref{fig:T}, are shown in Fig.~\ref{fig:FreeEnergies}.
As discussed in Sec.~\ref{sec:FiniteT}, the main effect of $K$ is to bring magnetic states higher up in energy at very low temperatures, so that they cross the solitonic branch at a finite temperature below the one at which it crosses the paramagnetic branch.
From these simple arguments, one also predicts that the soliton branch directly crosses the paramagnetic branch for all values of $K$ larger than the value $K_{max}$ for which $F_{sol}$, $F_{pol}$ and $F_{para}$ meet at a single point (here, $K_{max} \approx 0.36$).
These simple arguments thus allow us to rationalise the finite temperature results presented above. 

\section{Ordering of Velocities \label{sec:Velocities}}
%
\begin{figure}
    \centering
    \includegraphics[width = .49\columnwidth]{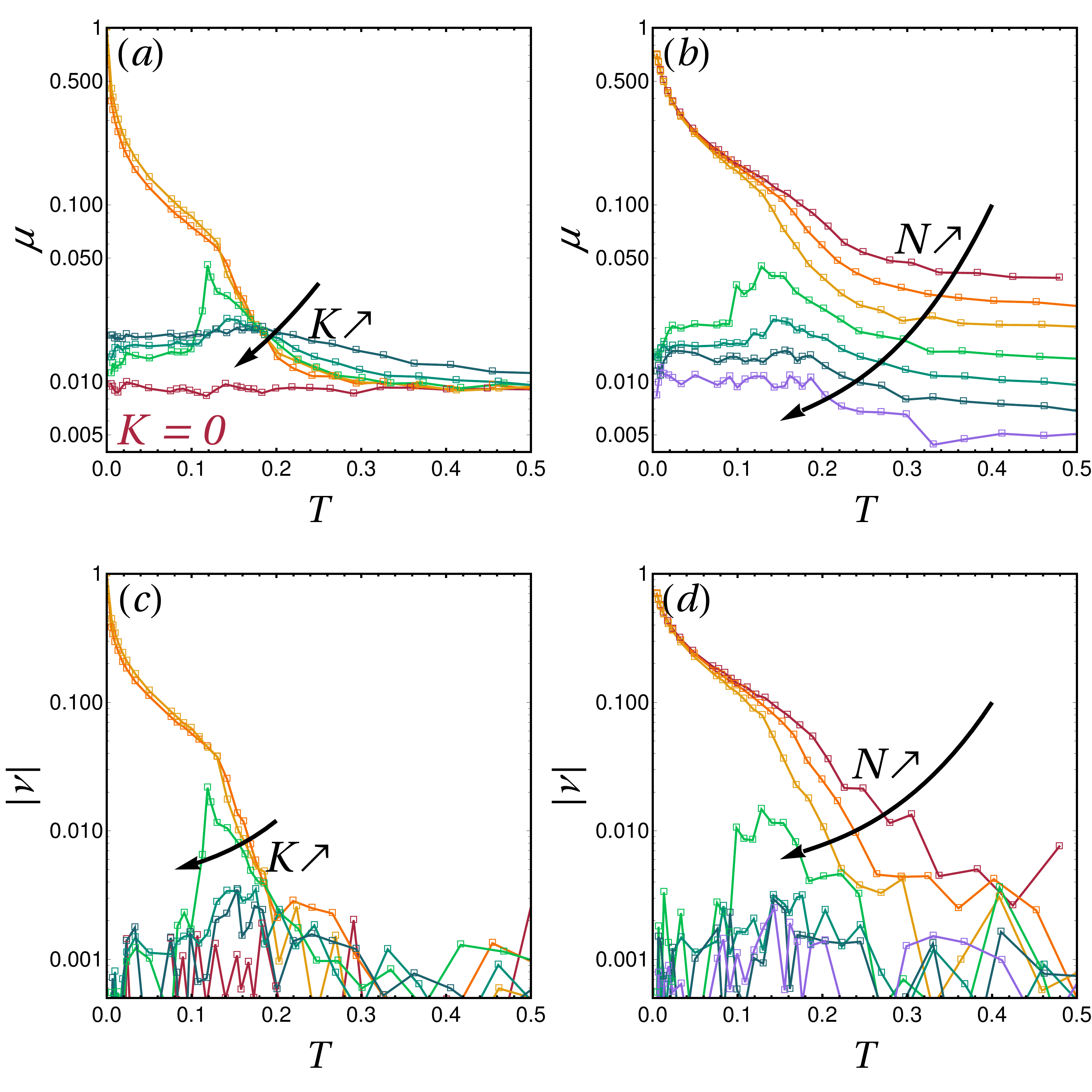}
    \includegraphics[width = .47\columnwidth]{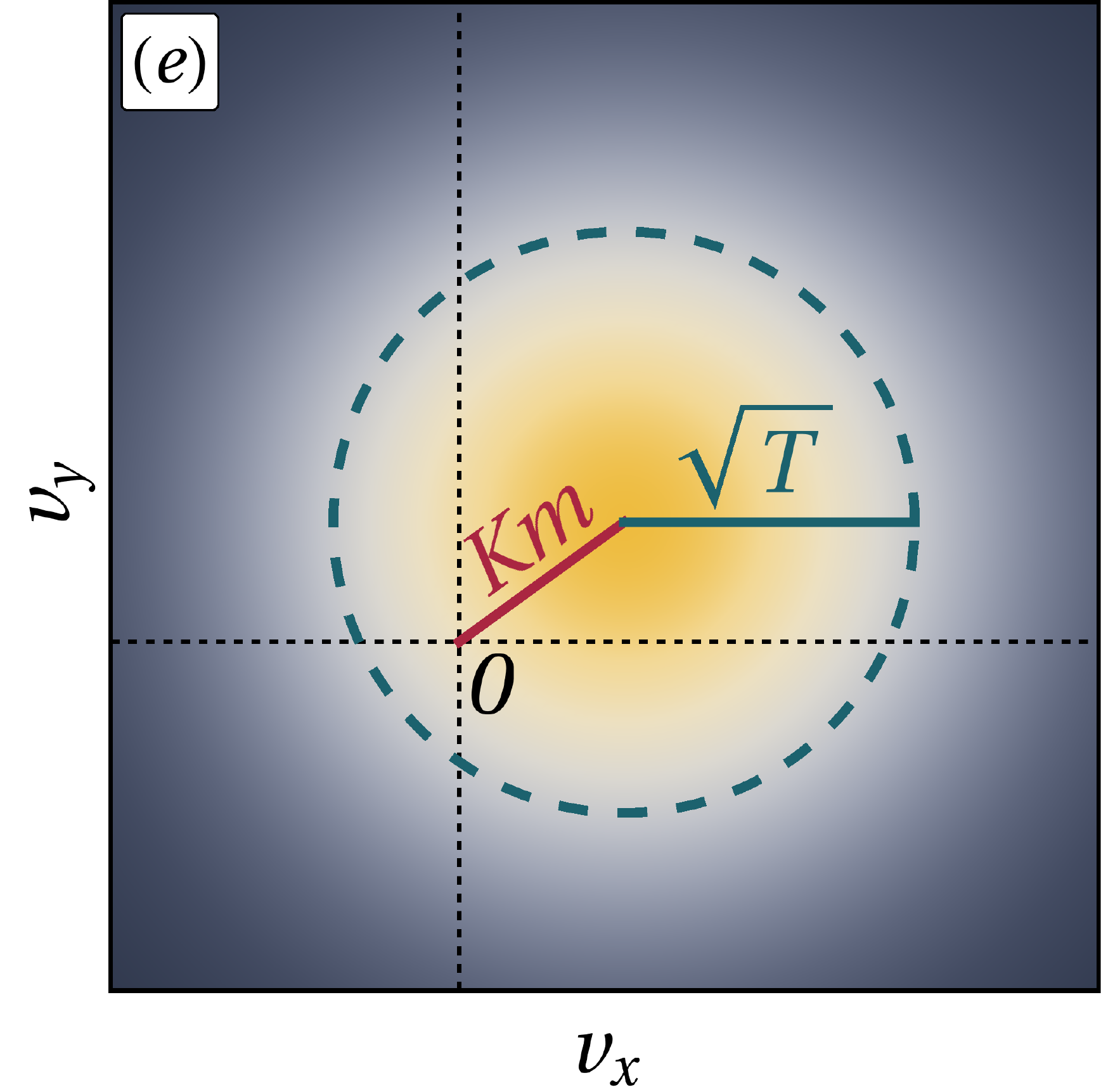} \\
    \includegraphics[width = .96\columnwidth]{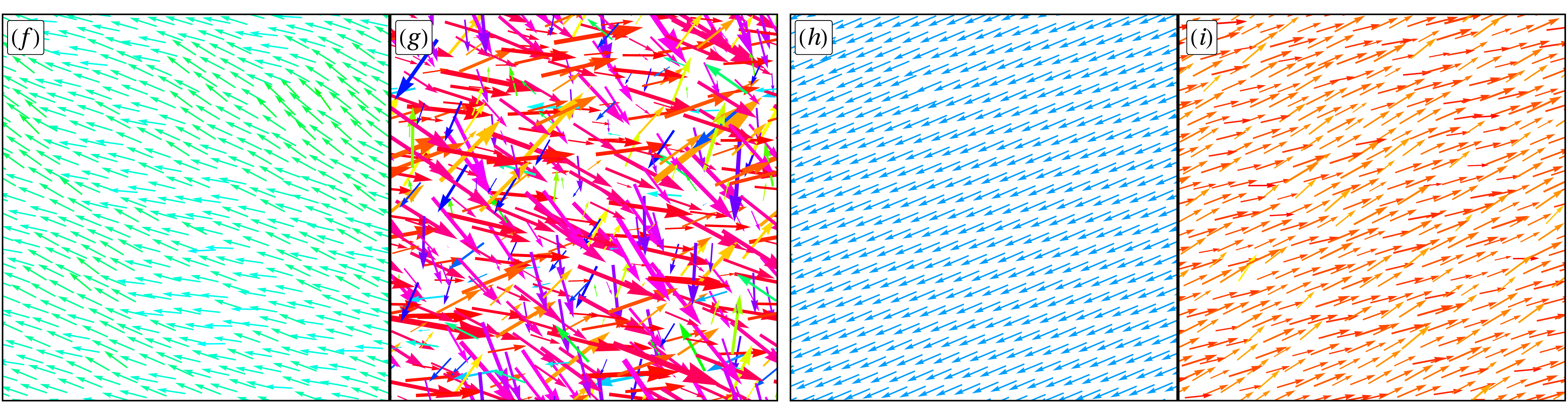}
    \caption{{\textbf{Ordering of velocities.}}
    $(a)$ Polarity of the velocities versus temperature, in log scale, for $N = 2048$ and varying $K = 0, 0.05, 0.06, 0.07, 0.1, 0.2$.
    $(b)$ Polarity versus temperature, this time fixing $K = 0.1$ and taking $N = 128, 256, 512, 1024, 2048, 4096, 8192$.
    $(c)$ Spin-velocity alignment versus temperature, in log scale, corresponding to $(a)$.
    $(d)$ Spin-velocity alignment versus temperature, in log scale, corresponding to $(b)$.
    The colours are the same as in Fig.~\ref{fig:T}.
    $(e)$ Sketch of the distribution of velocity vectors in magnetised phases, shown as a density plot, with values growing from grey to yellow. 
    The distribution is centered on $\bm{v}_G$, at a distance $K m$ from the origin (red line), and has a standard deviation of the order of $\sqrt{T}$ (blue circle).
    $(f)-(g)$ Snapshots of the magnetisations and velocities, respectively, for $N = 512$ and $K = 0.1$ at $T \approx 7  \cdot10^{-3}$.
    The colour of each vector represents its orientation, while its thickness is proportional to its length.
    $(h)-(i)$ Snapshots of the same system at $T \approx 3\cdot10^{-6}$.}
    \label{fig:velocities}
\end{figure}

We are yet to describe how velocities order in polar moving states.
Because of the conservation of the total momentum, any state with a macroscopic magnetisation also moves with a collective velocity $\bm{v}_G = - K \bm{m}$.
Therefore, one expects velocity vectors to align together (and to select the same speed) as the temperature is lowered.
It is thus interesting to define an order parameter that reflects the alignment properties of the velocities, 
\begin{align}
    \mu = \left\langle\left|\frac{1}{N}\sum\limits_{i=1}^{N} \bm{\hat{v}}_i \right|\right\rangle,
\end{align}
which reflects the alignment of unit velocities regardless of the velocity moduli, and is often called the polarity in the context of active matter~\cite{Cavagna2010}.
We furthermore define an observable that measures the alignment of unit velocities onto the spins,
\begin{align}
    \nu = \left\langle\frac{1}{N}\sum\limits_{i=1}^{N} \bm{\hat{v}}_i\cdot \bm{s}_i  \right\rangle.
\end{align}
The variations of $\mu$  and $\nu$ with temperature, obtained for increasing $K$ at fixed $N$ or increasing $N$ at fixed $K$ are shown in Fig.~\ref{fig:velocities}$(a)-(d)$.

For low values of $K$ and $N$, such that the system has a polar ground state, both $\mu$ and $\nu$ grow to $1$ at low temperatures.
However, these curves do not behave like the magnetisation: $\mu$ and $\nu$ become significantly larger than zero only at very low temperatures (note the logarithmic scale of the $\mu$ and $\nu$ axis).
This is even more striking when comparing the magnetisation and velocity fields at low temperatures, as illustrated by the snapshots in Fig.~\ref{fig:velocities}$(f)-(i)$.
In particular, in the snapshots~\ref{fig:velocities}$(f)-(g)$, we observe that even deep in the magnetised phase, where the system moves with a collective velocity, both the orientations (colours) and moduli (lengths and thicknesses) of the velocity arrows seem to vary a lot across the system.
This comparatively weak ordering of the velocities compared to that of the spins can be explained by focusing on the distribution of velocity vectors.
Rewriting the equipartition equation~(\ref{eq:EquipMom}) associated to momenta in terms of the velocities, one finds that
\begin{align}
    2 T &= \left\langle v_i^2 \right\rangle - \left\langle v_i \right\rangle^2 + 2 K \left( \left\langle \bm{v}_i\cdot\bm{s}_i \right\rangle - \left\langle \bm{v}_i \right\rangle \cdot \left\langle \bm{s}_i \right\rangle \right). \label{eq:EquipVel}
\end{align}
In particular, deep in the magnetised phase, one can assume that all spins are essentially aligned and equal to the (unit) magnetisation vector, $\bm{s}_i \approx \bm{m}$.
In this simple case, the spin-velocity terms in Eq.(~\ref{eq:EquipVel}) cancel out, so that the usual equipartition relation, $2 T = \left\langle v_i^2 \right\rangle - \left\langle v_i \right\rangle^2$, is recovered.
Consequently, at very low temperatures, the velocity vectors are essentially drawn from a Gaussian distribution centered at $\bm{v}_G = - K \bm{m}$, and with an isotropic variance $T$.
This distribution is sketched as a density plot in Fig.~\ref{fig:velocities}$(e)$, with values growing from gray to yellow.
From this sketch, it is immediately visible that, as long as the standard deviation of the velocity distribution is larger than $K$, meaning that $T \gtrsim K^2 $, there is a significant probability for velocities to be pointing in any direction other than that of $\bm{v}_G$.
Their moduli, however, are skewed towards larger values.
Hence the aspect of the snapshot in Fig.~\ref{fig:velocities}$(g)$, in which velocities display a low polar order, but are biased towards larger, red arrows.
In practice, for polar order to be visible, like in snapshot~\ref{fig:velocities}$(i)$, one needs to reach temperatures such that $T \ll K^2$, thus ensuring that the Gaussian distribution of the velocities does not take significant values close to $\bm{v} = \bm{0}$.
This constraint on the value of the temperature to observe polar order of the velocities is usually absent in models that feature collective motion.
The reason for this is two-fold.
First, in abstract models of self-propelled particles like the Vicsek model~\cite{Vicsek1995}, the speed of particles is often fixed to a nominal value, so that the distribution of the velocities is naturally off-centered and very sharp.
In that sense, a conservative setting makes it harder to observe a polar order of the velocities, since it imposes constraints on the width of the speed distribution.
Second, in order to stabilise the polar state, $K$ needs to be small enough, so that the velocity of the centre of mass is never that large and $\mu$ can only grow at very low temperatures.

For very large values of $K$ or $N$, such that the ground state is not polar and no ObD is observed at finite temperatures, $\mu$ remains low at all temperatures.
This is expected, since no collective velocity develops in this case.
Furthermore, $\nu$ also remains small, thus indicating that the velocities do not locally align on the spins either.
In other words, in vortex and soliton structures, the velocities do not draw solitons or vortices themselves, but simply remain disordered at low temperatures: velocity alignment is destroyed.
This is proof that there is no true microscopic source of velocity alignment.
At first sight, one would have expected the introduction of the self-alignment of the velocity on the spin in the Lagrangian to be responsible for a transfer of the ferromagnetic alignment of the spins to the velocities, even in vortex or soliton states. 
This is not the case because the alignment of the velocities is actually entirely driven by the constraint imposed by the momentum conservation: $\bm{v}_G = - K \bm{m} = 0$. 
This constraint is only global and promotes local alignment between the velocities in a very indirect way. 

Finally, for intermediate values of $K$ or $N$, however, a macroscopic magnetisation is observed at finite temperatures, following the ObD scenario described in Sec.~\ref{sec:ObD}.
As a consequence, a collective velocity is also observed at finite temperatures, and it is accompanied by a local maximum of both $\mu$ and $\nu$.
In other words, the ObD phenomenon saves collective motion from a total extinction caused by topological defects at finite temperatures in a finite range of the parameters.
Furthermore, in the range of $K$ and $N$ such that we observe switches between polar and soliton states, such as those described in Fig.~\ref{fig:Crossover}, the collective velocity itself also switches between low and high values.
This spontaneous intermittent motion is highly unusual in a conservative setting, as such behaviours are usually expected in systems with friction, like granulars,~\cite{Rao2008} or activity, such as \textit{E. Coli} bacteria that alternate between ``runs'', i.e. ballistic motion along straight lines, and ``tumbles'', i.e. pauses in the motion during which they re-orient themselves.~\cite{Berg1977}

\section{Discussion and Conclusion}
\label{sec:Conclusion}
We studied a $2d$ Hamiltonian model of interacting particles that carry continuous spins. 
The latter interact ferro-magnetically with their neighbours, and are locally coupled to their velocities.
The homogeneous phases exhibit a rich behaviour, including the existence of frustration-induced topological defect configurations as ground states, ObD-like transition, and collective motion. 
Topological defects appear as a consequence of a classical spin-orbit coupling, which is reminiscent of the so-called topological phase transitions observed in quantum models when a spin-orbit coupling is added~\cite{Hasan2010}, and could have deeper links with active matter, as parallels between quantum theories with a spin-orbit coupling and flocking have recently been proposed~\cite{Loewe2018}. 
In Appendix~\ref{app:GeoDim}, we show that our results are unchanged when moving from $d=2$ to $d=3$, or when changing the geometry of the boundary conditions. 
Our results can therefore be seen as robust.

From the point of view of magnetism, despite the fact that the model is at first sight a bit peculiar, one can think of it as an efficient dynamics to prepare topological defect configurations, for instance in dimension $d\geq 3$, in which case observing and describing topological excitations is notoriously challenging.~\cite{Ackerman2017} 
From the point of view of collective motion, the model seems limited in terms of the amplitude of the velocities it gives access to, especially for large systems. 
A solution would be to choose $K$ such that it scales with the size of the system, but this rather unphysical choice would also lead to a vanishingly small collective speed in the large $N$ limit.
One could also in principle think of other coupling scheme between the spins and velocities, and hope for stronger effects. 
However, we show in Appendix~\ref{app:SOShape} that the proposed spin-velocity coupling is, in fact, one of the only two ones that are compatible with well-defined Hamiltonian dynamics: adding coupling terms beyond the quadratic order in velocity, leads to a situation in which the Hamiltonian dynamics can not be obtained from the Legendre transform of the Lagrangian.~\cite{Chi2014} 
We have also checked that adding a quadratic coupling to the model with a linear coupling $K\neq 0$ leads to the same phenomenology. 
The present model is therefore the {\textit{only}} reasonable conservative model in which a spin-velocity coupling leads to collective motion. 
Yet, the case of a model with only a quadratic coupling but no linear coupling, $K = 0$, can be of interest to describe a situation in which a nematic ordering of the velocities onto the direction of the spin should be favoured, as in a recent model of Vicsek-like particles with velocity reversals.~\cite{Mahault2018}

Finally, in the context of assemblies of self-propelled particles, regular arrangements of vortices~\cite{Sumino2012,Nagai2015} and soliton-like structures~\cite{DeCamp2015,Putzig2016,Patelli2019} have been described in experiments, simulations, and microscopic models.
This could be an indication that such structures are generally less costly than homogeneous motion in models of collective motion, regardless of the presence or absence of activity.

\acknowledgements{We would like to warmly thank Michael Schindler for his great help on the numerical aspects of this paper. We would also like to thank \'{E}ric Bertin, Jorge Kurchan, Jean-No\"{e}l Fuchs, and Pascal Viot for insightful discussions. Leticia F. Cugliandolo and Marco Tarzia are members of the \textit{Institut Universitaire de France}.
}

\bibliography{Bibtex-SpinOrbit}

\appendix
\section{Numerical methods\label{app:Num}}

In this Appendix we motivate the choice of Molecular Dynamics as an efficient simulation method 
to study the model on which we focus. 
In particular, we give details on the precise strategy 
leading to the figures shown in the main text. Moreover, 
we describe the additional difficulties encountered when simulating 
the same system in $3d$ but with continuous 
$3d$ (Heisenberg) spins, as described in App.~\ref{app:GeoDim}.

\subsection{Choice of Molecular Dynamics}

Simulating a system with a spin-velocity coupling presents several unusual aspects, both fundamental and practical.
On the fundamental side, the fact that the system is not Galilean invariant implies that the Gibbs measure should take its full, 
frame-dependent form~\cite{Diu1996,Attard2002,Bore2016}
\begin{align}
    \mathcal{P}_{\beta,\bm{v}_G}\left[\mathcal{C}\right] &\propto e^{- \beta\left(\mathcal{H}\left[\mathcal{C}\right] - \bm{v}_G\cdot\bm{P}\left[\mathcal{C}\right] \right)}
    \; .
\end{align}
$\beta$ and $\bm{v}_G$ are, respectively, the inverse temperature and the velocity of the 
centre of mass, 
and are both fixed through an external bath in the canonical ensemble. 
$\mathcal{H}$ and $\bm{P}$ are the 
Hamiltonian and the total momentum of the configuration $\mathcal{C}$, respectively. 
Therefore, because of the non-Galilean character of the model, in the canonical ensemble, the 
variables 
imposed by the bath are the temperature and the velocity.
As a consequence, usual canonical ensemble sampling via a Monte Carlo-like scheme is 
not expected to be the most 
interesting setting since the aim of the model is to demonstrate the {\it emergence} of a spontaneous $\bm{v}_G$ as a 
response to a lowering of the temperature or energy.
It seems more natural, instead, to impose a total momentum $\bm{P}$.
A possibility would then be to impose a temperature and a total momentum, that is to say, to work in an intermediary 
ensemble that lies between the microcanonical (conserved energy and momentum) and the canonical 
(conserved temperature and momentum).
However, little is known on the peculiarities of non-Galilean statistical mechanics, so that we choose 
the simpler setting of the microcanonical ensemble, that also enables us to see the actual dynamics of 
the system, for instance, the propagation of solitons.

In practice, we therefore integrate the Hamiltonian equations of motion,
\begin{align}
    \dot{\bm{r}}_i &= \bm{p}_i - K \bm{s}_i, \\
    \dot{\bm{p}}_i &= \sum\limits_{k (\neq i)} \left(\frac{\partial J (r_{ik})}{\partial \bm{r}_i} \cos\theta_{ik} - \frac{\partial U (r_{ik})}{\partial \bm{r}_i} \right), \\
    \dot{\theta}_i &= \omega_i, \\
    \dot{\omega}_i &= K \bm{p}_i \cdot \bm{s}_{\perp, i} + \sum\limits_{k (\neq i)} 
    J(r_{ik})\sin\theta_{ik}, \label{eq:HEoM}
\end{align}
where $\hat{\bm{s}}_\perp$ is the unit vector obtained by rotating the spin of an anticlockwise $\pi/2$ angle, and we choose a total momentum $\bm{P} = \bm{0}$. 
A technical issue, however, still remains.
In Molecular Dynamics, the equations of motion are usually integrated using numerical tricks 
like the Verlet algorithm~\cite{Verlet1967}, that ensure ``symplecticity", 
meaning that the momentum and the energy are rigorously conserved.
These tricks, however, usually rely on updating the two sets of conjugate Hamiltonian variables at different times.
Here, these methods are inapplicable due to the coupling between the velocity and the spin.
Instead, we simply reproduced the simulation strategy used in Ref.[~\onlinecite{Bore2016}], namely, a fourth-order Runge-Kutta 
scheme that conserves the total momentum only approximately.
However, the latter always remains small and fluctuates with a typical magnitude of $10^{-17}-10^{-14}$ in simulations, and is therefore much smaller than any other observable we considered.

\subsection{Simulation strategy}

In the main text, we simulate the dynamics starting from random states with uniformly distributed 
$\left\{\bm{r}_i, \theta_i \right\}_{i=1..N}$ and $\left\{\bm{p}_i, \omega_i \right\}_{i=1..N}$ drawn 
from centered, reduced Gaussian distributions. 
Such initial states were placed into a square box with periodic boundary conditions and, after giving some time for the dynamics to settle in, are subjected either to a numerical annealing or to a high-rate quench. 
These procedures are implemented as follows.

Numerical annealing is performed by multiplying all rotational velocities by $\lambda_{A} = 0.9999$ every 100 time units in our adimensional variable, with an integration time step equal to $\delta t = 10^{-3}$ in the same units. 
This method enables us to reach low-energy states which, if the cooling is slow enough, should be equilibrium states.

Quenches are carried out by multiplying all rotational velocities and momentum components 
by $\lambda_{Q} = 0.10$ once, at some initial time. 
This method violently takes the system away from equilibrium, thereby enabling us to study the subsequent equilibration dynamics. 

Whenever curves represent observable mean values, we mean that we averaged the results over $10^2-10^3$ independent 
configurations obtained by sampling initial conditions and recording states at times that are far enough 
from each other that the corresponding configurations can be considered to be independent.

\subsection{Variant: $3d$ case}

In order to simulate the system in 3 dimensions, we also used the Runge-Kutta Molecular Dynamics method. In this higher dimensional case the proper Hamiltonian dynamics are given by
\begin{eqnarray}
 \bm{\dot{r_i}}      &=& \bm{p_i} - K_1 \bm{s_i} \; , \\
 \bm{\dot{p_i}}      &=& \frac{1}{2}\sum\limits_{k\neq i} \left(j'(r_{ik})\bm{s_i}\cdot\bm{s_k} - u'(r_{ik})\right)\bm{\hat{r}_i} \; , \\
 \dot{\theta}_i      &=& p_{\theta_i} \;, \\
 \dot{p}_{\theta_i}  &=& \frac{p_{\phi_i}^2 cos\theta_i}{\sin^3\theta_i} -\frac{\partial H_{int}}{\partial \theta_i} \; , \\
  \dot{\phi}_i       &=& \frac{p_{\phi_i}}{\sin^2\theta_i} \; , \\
 \dot{p}_{\phi_i}    &=& -\frac{\partial H_{int}}{\partial \phi_i} 
 \; , 
\end{eqnarray}
where we called $H_{int}$ the interaction part of the Hamiltonian. These equations pose a major problem to their 
numerical integration, as divergences at the poles appear in the evolution equations for $\phi_i$ and $p_{\theta_i}$, at 
each time step. This problem, which is usual for simulations of rotors on spheres, is here rather tricky to 
solve~\cite{Vest2014}, and a variety of strategies can be tested. In our case, we chose to write modified evolution 
equations in which we consider spins to be arbitrary Cartesian vectors, but that should (within numerical accuracy) 
conserve the spin normalisation.
The idea, adapted from previous works on Heisenberg spins~\cite{Nobre2003} and dynamics in a spherical geometry~\cite{Vest2014} consists in writing
\begin{eqnarray*}
 H \equiv \sum\limits_{i=1}^N\left[ \frac{\bm{{p_i}}^2}{2} + \frac{\bm{L_i}^2}{2} -  K_1 \bm{p_i}\cdot\bm{s_i} - \frac{1}{2}\sum\limits_{k(\neq i)} \left(j(r_{ik})\bm{s_i}\cdot\bm{s_k} - u(r_{ik})\right) \right]
 \; , 
\end{eqnarray*}
where $\bm{L_i} = \bm{s_i}\wedge\bm{p{s_i}}$ is the usual angular momentum of the  spins. 
For unit spins, $L_i^2 = \bm{p_{s_i}}^2 = \bm{\dot{s}_i}^2$, as their time derivative should be perpendicular to them. We then write the evolution equations:
\begin{eqnarray}
 \bm{\dot{s}_i} &=& \bm{L_i} \wedge  \bm{s_i} 
 \; , \\
 \bm{\dot{L}_i} &=& \bm{s_i}\wedge\left( - \frac{\partial H}{\partial \bm{{s_i}}} \right)
 \; . 
\end{eqnarray}
These equations are natural consequences of the definition of $\bm{L}_i$ with respect to $\bm{p_{s_i}}$, the actual canonical momentum associated to the spins. The advantage of this way of writing the evolution is that, 
even though it is not strictly symplectic in 3 dimensions, it does conserve the 
modulus of the spins up to numerical errors.

\section{The spin-velocity coupling\label{app:SOShape}}

\subsection{Breakdown of the Lagrangian-to-Hamiltonian transformation}

As mentioned in the main text, it is tempting to explore various forms of 
the spin-velocity coupling, in the hope that they may influence the phenomenology of the ground states in different ways.

A first remark is that due to the rotational invariance of the spins and velocities, 
any linear spin-velocity coupling with a shape
\begin{align}
    \mathcal{L}_{sv} &= K \bm{v_i}\cdot\mathcal{R}_\varphi\left[ \bm{s_i} \right]
    \; ,
\end{align}
where $\bm{v}_i = \dot{\bm{r}}_i$ is a short-hand notation for the velocity of the particle $i$, 
and $\mathcal{R}_\varphi$ is a rotation by an angle $\varphi$, would be equivalent to the definition we used.
The momentum  conservation would now impose a fixed angle determined by $\varphi$ between 
the velocities and spins at low energies.

Another possibility is to add an additional non-linear spin-velocity term to the Lagrangian.
By doing so, one could hope to find an additional velocity dependence in the Hamiltonian, 
and possibly a reduction of the cost of kinetic energy.
Let us first discuss the term that is perhaps the most natural to add, a quadratic coupling.
The spin-velocity interaction terms in the Lagrangian would then be
\begin{align}
     \mathcal{L}_{sv} &= \sum\limits_{i = 1}^{N} K_1  \bm{v_i}\cdot \bm{s_i} -  \sum\limits_{i = 1}^{N} K_2  \left(\bm{v_i}\cdot \bm{s_i}\right)^2
     \; ,
\end{align}
where we called $K_1$ the coupling constant of the linear term and $K_2$ the one in front of the 
quadratic coupling for clarity.
This coupling looks interesting, as for positive values of $K_2$ it decreases the kinetic cost of 
velocities aligned with the spins.
With this coupling, the canonical linear momentum reads
\begin{align}
    \bm{p}_i = \dot{\bm{r}}_i + K_1 \bm{s}_i - 2 K_2 \left(\dot{\bm{r}}_i \cdot \bm{s}_i \right) \bm{s}_i
    \; . 
\end{align}
An issue with this expression is that, as it is not linear in $\dot{\bm{r}}_i$, 
it is not easy to invert it and write explicitly the velocity in terms of $\bm{p}_i$ 
and $\bm{s_i}$ only. Moreover, 
this fact poses a problem in the Hamiltonian formalism,
\begin{align}
    \mathcal{H} &= \sum\limits_{i = 1}^{N} \left[ \bm{v}_i \cdot \bm{p}_i + \omega_i^2 \right] - \mathcal{L} \nonumber \\
                &= \sum\limits_{i = 1}^{N} \left[ \frac{1}{2}\omega_i^2 + \bm{v}_i \cdot \bm{p}_i - \frac{1}{2}v_i^2 - K_1 \bm{v}_i\cdot\bm{s}_i + K_2 \left( \bm{v}_i\cdot\bm{s}_i\right)^2  \right] - \mathcal{L}_{int}
                \; ,
\end{align}
where $\mathcal{L}_{int}$ is the interaction part of the Lagrangian.
The Hamiltonian above can be easily  rewritten in terms of 
the $\bm{v}_i$'s and $\bm{s}_i$'s only, and reads
\begin{align}
    \mathcal{H} &= \sum\limits_{i = 1}^{N} \left[ \frac{1}{2}\omega_i^2 + \frac{1}{2}v_i^2 - K_2 \left( \bm{v}_i\cdot\bm{s}_i\right)^2  \right] - \mathcal{L}_{int}
    \; .
\end{align}
As hoped when defining the quadratic term in the Lagrangian, this term decreases the cost of the kinetic energy, even in the Hamiltonian. However, 
one can clearly see that problems arise for $K_2 \geq 1/2$, since in this range of values 
the energy can be made arbitrarily low for $v_i \to \infty$.
In fact, in this regime, the dynamics themselves are ill-defined, because the Lagrangian is no longer convex with respect 
to the velocities, and this leads to unphysical results like 
infinite accelerations or velocities even 
for a single particle system.
While there are ways to exploit such theories using path integral formulations~\cite{Chi2014} as well as interpretations 
of their peculiarities in quantum models, we will here restrict ourselves to $K_2 < 1/2$.
Note that the convexity condition imposed by $K_2<1/2$ 
would lead to similar restrictions for other non-linear couplings, with several 
forbidden intervals for their corresponding coupling constants.

Another issue 
is to rewrite the Hamiltonian in terms of its canonical variables.
It is here more complicated than usual, as the only way to do so is to project the canonical momentum 
onto the vectors of interest, $\bm{v}_i$, $\bm{s}_i$, and $\bm{p}_i$. After some algebra we obtain
\begin{align}
    \bm{v}_i\cdot \bm{s}_i &= \frac{\bm{p}_i\cdot \bm{s}_i - K_1}{1 - 2 K_2}, \label{eq:vs}\\
    \bm{v}_i\cdot \bm{p}_i &= v_i^2 + K_1 \bm{v}_i\cdot \bm{s}_i - 2 K_2 \left(\bm{v}_i\cdot \bm{s}_i\right)^2, \\
        v_i^2              &= p_i^2 + A \bm{p}_i \cdot \bm{s}_i + B  \left(\bm{p}_i\cdot \bm{s}_i\right)^2 + C
        \; ,
\end{align}
where we defined the constants
\begin{align}
    A &= - \frac{2 K_1}{\left(1 - 2 K_2\right)^2} \; , \qquad\qquad
    B = \frac{4 K_2 (1- K_2)}{\left(1 - 2 K_2\right)^2} \; , \qquad\qquad
    C = \frac{K_1^2}{\left(1 - 2 K_2\right)^2} \; . 
\end{align}
Using these projections, after some more algebra, we can finally rewrite the Hamiltonian as a function of the 
canonical variables only
\begin{align}
    \mathcal{H} &= \sum\limits_{i = 1}^{N} \left[ \frac{1}{2}\omega_i^2 + \frac{1}{2}p_i^2 - \tilde{K}_1 \bm{p}_i\cdot\bm{s}_i + \tilde{K}_2 \left(\bm{p}_i\cdot \bm{s}_i\right)^2  \right] - \mathcal{L}_{int}
    \; , \label{eq:K2Ham}
\end{align}
where the Hamiltonian coupling constants $\tilde{K}_1$ and $\tilde{K}_2$ are, interestingly, 
defined differently from the Lagrangian ones,
\begin{align}
    \tilde{K}_1 &= \frac{K_1}{1 - 2 K_2} 
    \; , \qquad\qquad
    \tilde{K}_2 = \frac{K_2}{1 - 2 K_2}
    \; ,
\end{align}
and where we discarded a constant term $E_0$ defined as
\begin{align}
    E_0 &= \frac{K_1^2}{2(1 - 2 K_2)}
    \; .
\end{align}
It is a priori much more difficult and maybe even impossible to perform this inversion for any other non-linear couplings, as the vector projections used here no longer yield linear equations between the terms of interest.
In particular, as tempting as a coupling of the form $\hat{\bm{v}}_i\cdot \bm{s}_i$, where $\hat{\bm{v}}_i = \bm{v}_i / v_i$ looks, it yields a system of equations that do not allow to write a Hamiltonian in terms of the corresponding canonical momentum.
Also note that higher order polynomials in $\bm{v}_i\cdot \bm{s}_i$ yield, instead of Eq.~(\ref{eq:vs}), polynomial equations in $\bm{v}_i\cdot \bm{s}_i$ with degree higher or equal to two, so that several roots exist.

Therefore, the only two reasonable couplings one can imagine to get unequivocal Hamiltonian dynamics with a spin-velocity coupling are the linear and quadratic couplings.
An important message is that velocity terms in a conservative model are very constrained, so that it is in general necessary to take the system out of equilibrium by breaking energy or momentum conservation if one wants to act arbitrarily on velocities.
Any other non-linear term will lead either to convexity
or to $\bm{p} \leftrightarrow \bm{v}$ inversion issues.

Regarding the quadratic coupling introduced in this part, it does not seem to make the system simpler nor to lead to new physics relevant to collective motion for $K_1 > 0$.
Indeed, if we reproduce the logic of the first study of the model with $K_1$ only, and consider only cases with $\bm{P} = \bm{0}$, we expect collective motion only for magnetised states.
In these states, assuming that we only consider states with $\bm{s}_i = \bm{m}, \forall i$ the momentum conservation constraint is now $\bm{v}_G = - \tilde{K}_1 \bm{m}$, so that the velocity of the centre of mass will just be rescaled in the same way as the coupling constants in the Hamiltonian. Therefore, we 
roughly expect to observe the same states as in the case $K_2 = 0$.
Further proof that we do not expect any huge qualitative change when adding $K_2$ can be obtained by computing the partition function and observables in a mean-field approximation.
The full calculation is reported in Sec.~\ref{app:K2MF}, and shows that the results are essentially the same as those obtained with $K_1$ only~\cite{Bore2016}.
In particular, the magnetisation curves remain almost unaltered by the addition of $K_2$.

The case $K_1 = 0, K_2 > 0$, however, might be interesting to study on its own, as it favours nematic order between the velocity and the spins. This is reminiscent of a recent model of active matter in which velocities are allowed to switch between alignment and anti-alignment with an internal spin degree of freedom.~\cite{Mahault2018}
It however looks like a different problem than the one of collective motion, as we do not expect any spontaneous flow of particles to arise in the case $K_1 = 0$, whatever the value of $K_2$.
In light of these results, we are confident that the model with a single $K_1$ should contain all the physics linked to the addition of a conservative, and polar spin-velocity coupling.

\subsection{Mean-field calculation with $K_2$\label{app:K2MF}}

We report in this Section the mean-field study of the Hamiltonian written in Eq.~(\ref{eq:K2Ham}), 
\begin{align}
    \mathcal{H} &= \sum\limits_{i = 1}^{N} \left[ \frac{1}{2}\omega_i^2 + \frac{1}{2}p_i^2 - \tilde{K}_1 \bm{p}_i\cdot\bm{s}_i + \tilde{K}_2 \left(\bm{p}_i\cdot \bm{s}_i\right)^2  \right] - \mathcal{L}_{int}
    \; , 
\end{align}
where 
\begin{align}
    \mathcal{L}_{int} &= \sum\limits_{i \neq j} \left[J(r_{ij}) \bm{s}_i\cdot\bm{s}_j - U(r_{ij}) \right]
    \; .
\end{align}

In order to get a mean-field description, we follow the same strategy as in Ref.~\cite{Bore2016} where 
case $K_2 = 0$ was dealt with in detail: we set 
$U = 0$ and we take the fully-connected limit for the ferromagnetic coupling by setting
\begin{align}
    J(r_{ij}) &= \frac{J}{N}
    \; ,
\end{align}
with $J$ a constant.
Then, discarding a constant term, the Hamiltonian can be rewritten as
\begin{equation}
\mathcal{H} = \sum\limits_{i=1}^N \left[ \frac{{\omega_i}^2}{2} + \frac{{p_i}^2}{2} - \tilde{K}_1 \left(\bm{p_i}\cdot\bm{s_i}\right) + \tilde{K}_2 \left(\bm{p_i}\cdot\bm{s_i}\right)^2 - \frac{J}{2} \bm{s_i}\cdot \bm{m}\right]
\end{equation}
and, recalling that the Gibbs measure is altered by the fact that this system is not Galilean invariant, the mean-field partition function reads
\begin{equation}
 Z \equiv \int d^{2N}\bm{R} \, d^{2N}\bm{\mathcal{P}} \, d^N \bm{\Theta} \, d^N \bm{\Omega} \, \exp\left[- \beta \left( \mathcal{H} - \bm{v}_G \cdot \bm{P} \right) \right]
 \; ,
\end{equation}
where we used the short-hand notations $d^{2N}\bm{R} = \prod_{i = 1}^N d^2 \bm{r}_i$, $d^{2N}\bm{P} = \prod_{i = 1}^N d^2 \bm{p}_i$, $d^N \bm{\Theta} = \prod_{i = 1}^N d \theta_i$, and $d^N \bm{\Omega} = \prod_{i = 1}^N d \omega_i$.
The integrals over the positions and angular velocities factorise and can be easily computed, yielding
\begin{align}
 Z &= \left(\frac{2\pi}{\beta}\right)^{N/2} V^N \int d^N \bm{\Theta}  \;  \prod\limits_{i=1}^N Z_{p_i}(\theta_i) \exp\left(  \frac{\beta J}{2} \bm{s_i}\cdot\bm{m}   \right) 
 \; , \\
 Z_{p_i}(\theta_i) &\equiv \int d^2 \bm{p_i} \; \exp\left[ - \beta \left( \frac{{p_i}^2}{2} - \bm{p_i}\cdot\left( \bm{v}_G + \tilde{K}_1 \bm{s_i}\right)  + \tilde{K}_2 \left(\bm{p_i}\cdot \bm{s_i} \right)^2  \right)\right]
 \; . 
\end{align}
Calling $s_{i_{x,y}}$ the projections of spins onto the $x,y$ direction, the exponent in the last expression can be rewritten in a more compact form by using a vector notation:
\begin{eqnarray}
 Z_{p_i}(\theta_i) &\equiv& \int d^2 \bm{p_i} \; \exp\left( - \frac{1}{2}{^t}\bm{p} A \bm{p}    + {^t}\bm{B} \cdot \bm{p} \right)
 \end{eqnarray}
 with 
 \begin{eqnarray}
 A &\equiv& 2 \beta \; 
 \begin{pmatrix} 
 \frac{1}{2} + \tilde{K}_2 {s_{i_x}}^2  & {s_{i_x}}{s_{i_y}}\tilde{K}_2 
 \vspace{0.15cm}
 \\ 
 {s_{i_x}}{s_{i_y}} \tilde{K}_2 &   \frac{1}{2} + \tilde{K}_2 {s_{i_y}}^2     
 \end{pmatrix} 
 \qquad\qquad
 \mbox{and}
 \qquad\qquad
 {^t}\bm{B} \equiv \beta  \left( \bm{v}_G + \tilde{K}_1 \bm{s_i} \right)
 \; ,
\end{eqnarray}
and the integration over $\bm{p}_i$ readily calculated to obtain
\begin{equation}
 Z_{p_i}(\theta_i) = \sqrt{\frac{4 \pi^2}{\mathrm{det} A}} \; \exp\left( \frac{1}{2} {^t}\bm{B} A^{-1} \bm{B} \right)
 \; . 
\end{equation}
Moreover, using the fact that $\bm{s_i}$ is a unit vector, one finds
\begin{align}
 \mathrm{det}A &= \beta^2\left( 1 + 2 \tilde{K}_2\right), \\
 \frac{1}{2}{^t}\bm{B} A^{-1} \bm{B} &= \frac{1}{2 \, \mathrm{det}A}\left(B_1^2 A_{22} + B_2^2 A_{11} - 2 A_{21} B_1 B_2 \right) 
 \; , \nonumber \\
                          &= \frac{ \beta}{1+2\tilde{K}_2} \left[\frac{{\tilde{K}_1}^2 + v_G^2}{2} + \tilde{K}_1 \bm{v}_G\cdot\bm{s_i} - 2 \tilde{K}_2 v_{G,x} s_x v_{G,y} s_y + \tilde{K}_2 \left( s_x^2 v_{G,y}^2 + s_y^2 v_{G,x}^2 \right) \right].
\end{align}
Note that in the case $K_2 > 1/2$, that was already flagged as non-physical right from the definition of the quadratic term, one finds that det$A$ becomes negative; and in the limit $K_2 \rightarrow 1/2$, 
this determinant goes through an infinite value. 
The last results can be used to write
\begin{equation}
 Z_{p_i}(\theta_i) = \frac{2\pi}{\beta \sqrt{1 + 2 \tilde{K}_2}} \exp\left\{ \frac{\beta}{1+2\tilde{K}_2}  \left[ \frac{{\tilde{K}_1}^2 +v_G^2}{2} + \tilde{K}_1 \bm{v}_G\cdot\bm{s_i} - 2 \tilde{K}_2 v_{G,x} s_x v_{G,y} s_y + \tilde{K}_2 \left( s_x^2 v_{G,y}^2 + s_y^2 v_{G,x}^2 \right) \right] \right\}
 \; .
\end{equation}
Then, we notice a number of identities:
\begin{eqnarray}
 && 2v_{G,x} v_{G,y} s_x s_y - s_x^2 v_{G,y}^2 - v_{G,x}^2 s_y^2 = \left(\bm{v}_G\cdot\bm{s} \right)^2 - v_G^2 \; ,  \\
 && \frac{1}{1 + 2 \tilde{K}_2} = 1 - 2 K_2 \; , \qquad\qquad
 \frac{\tilde{K}_1}{1+2\tilde{K}_2} = K_1 \; , \qquad\qquad
 \frac{\tilde{K}_2}{1+2\tilde{K}_2} =  K_2 
 \; , 
\end{eqnarray}
and we define $Z_\theta$ through
\begin{equation}
    Z_\theta \equiv  \int d^N \bm{\Theta}  \;  \prod\limits_{i=1}^N Z_{p_i}(\theta_i) \exp\left(  \frac{\beta J}{2} \bm{s_i}\cdot\bm{m}   \right),
\end{equation}
so that we can rewrite
\begin{eqnarray}
 Z_\theta &=& \left(\frac{2 \pi \sqrt{1 - 2K_2}}{\beta }\right)^N \exp\left[ \frac{\beta N}{2} \left( \frac{ K_1^2}{1 -2K_2} + v_G^2(1 - 2 K_2)\right)\right] \mathcal{I} 
 \end{eqnarray}
 with 
 \begin{eqnarray}
 \mathcal{I} &\equiv& \int d^N\bm{\Theta} \exp\left[ \beta N \bm{m}\left( \bm{h} + \frac{J N}{2}\bm{m} + K_1 \bm{v}_G \right) - \beta N K_2 v_G^2 g(\bm{\Theta})\right] 
 \; , \\
 g(\bm{\Theta}) &\equiv& \frac{1}{N}\sum\limits_{i=1}^N \cos^2(\theta_i - \theta_\alpha) - 1
 \; .
\end{eqnarray}
It is now useful to apply a Hubbard-Stratonovich transformation,
\begin{equation}
 \exp\left( \frac{\beta J}{2} N^2 m^2 \right) = \frac{\beta N}{2\pi} \int d^2\bm{u} \; \exp\left[ - \beta N \left( \frac{u^2}{2} - J N \bm{u}\cdot\bm{m}\right) \right]
 \; , 
\end{equation}
and the definition $\bm{\gamma}(\bm{u}) \equiv j N \bm{u} + \bm{h} + K_1 \bm{v}_G$, to derive a more 
convenient expression for $\mathcal{I}$
\begin{eqnarray}
 \mathcal{I} &=& \frac{\beta N}{2 \pi} \exp\left[ \beta N K_2 v_G^2 \right] \int d^2 \bm{u} \exp\left[ - \beta N \frac{u^2}{2} \right] \left[J(\bm{u}) \right]^N 
 \end{eqnarray}
 with 
 \begin{eqnarray}
 J(\bm{u}) &\equiv& \int\limits_0^{2\pi} d\theta \exp\left[ - \beta K_2 v_G^2 \cos^2\left( \theta - \theta_\alpha \right) + \beta \gamma \cos\left( \theta - \theta_\gamma\right) \right].
\end{eqnarray}
Therefore, the full partition function reads
\begin{eqnarray}
 Z &=& V^N \left(\frac{2\pi}{\beta}\right)^{N/2} \frac{\beta N}{2\pi} \left(\frac{2\pi\sqrt{1 - 2 K_2}}{\beta }\right)^N  \exp\left[\beta N \left( \frac{K_1^2}{2(1 - 2 K_2)} + \frac{v_G^2}{2}\right) \right] \int d^2\bm{u} \; e^{-\beta N \frac{u^2}{2}} J(\bm{u})^N 
 \nonumber\\
   &=&  \left(V \sqrt{1 - 2 K_2}\right)^N \left( \frac{2\pi}{\beta} \right)^{\frac{3N}{2}-1} N \exp\left[\frac{\beta N}{2} \left(\frac{K_1^2}{1 - 2 K_2} + v_G^2  \right)\right] \int d^2\bm{u} \; e^{-\beta N \frac{u^2}{2}} J(\bm{u})^N
   \; . 
   \end{eqnarray}
In order to compute $J(\bm{u})$, we need to evaluate the integral
\begin{eqnarray}
  f(\bm{a},b) = \int\limits_{-\pi}^{\pi} d\theta \exp[a \cos\theta + b \cos^2(\theta + \theta_a)]
  \; .
\end{eqnarray}
With this aim, we use the identities~\cite{Abramowitz1972}:
\begin{eqnarray}
 \exp\left[ b \cos^2 (\theta + \theta_a)  \right] &=& \sum\limits_{n = 0}^\infty  \frac{(b/4)^n}{n!}\left[ \begin{pmatrix} 2n \\ n \end{pmatrix} + 2 \sum\limits_{k=0}^{n-1} \begin{pmatrix} 2n \\ k \end{pmatrix} \cos\left(2 \left( n - k \right) \left(\theta + \theta_a\right)\right) \right] 
 \; , \\
 \cos\left(2 \left( n - k \right) \left(\theta + \theta_a\right)\right) &=& \cos\left(2 \left( n - k \right)\theta\right) \cos\left(2 \left( n - k \right)\theta_a\right) - \sin\left(2 \left( n - k \right)\theta\right)\sin\left(2 \left( n - k \right)\theta_a\right)
 \; .
\end{eqnarray}
\begin{figure}\centering
 \includegraphics[width=.45\textwidth]{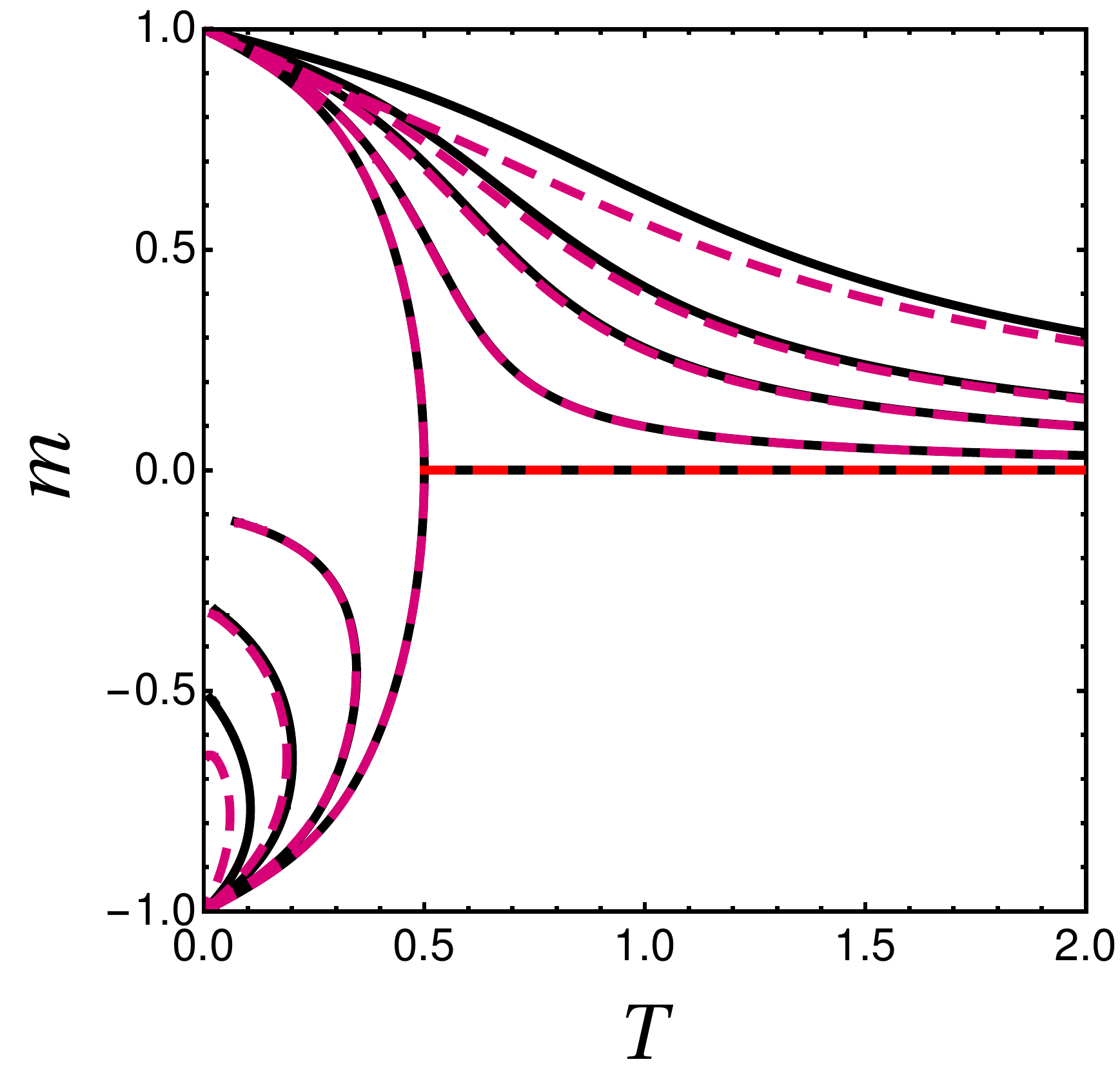}
\caption{The magnetisation density at the saddle point.  $K_1$ is taken equal to 1 and we plotted several values of the modulus of the velocity of the centre of mass $v_G = 0, 0.1, 0.3, 0.5, 1$. The red curves are obtained for $K_2 = 0.49$. The black curves are a reference, and represent the magnetisation in the case $K_2 = 0$.
\label{fig:K2m}}
\end{figure}
Then, integrating out odd terms that involve sines, and after some algebra, we find an exact expression for $f$:
\begin{eqnarray}
          f(\bm{a},b)      &=& 2\pi e^{\frac{b}{2}} \left(  I_0(a) I_0(\frac{b}{2}) + 2 \sum\limits_{j = 1}^\infty \cos(2 j \theta_a) I_j(\frac{b}{2}) I_{2j}(a) \right)
          \; . 
\end{eqnarray}
Taking advantage of the fact that $J(\bm{u}) =  \left[f(\beta \bm{\gamma},- \beta K_2 v_G^2)\right]^N$, 
we can now move on to the computation of the integral over $\bm{u}$:
\begin{equation}
\mathcal{J} = \int_{\mathbb{R}^2}d^2 \bm{u} \;  
\exp \left[ - \beta N \left( \frac{u^2}{2} - \frac{1}{\beta} \log\left[ f(\beta \bm{\gamma},- \beta K_2 v_G^2) \right] \right)  \right]
\; . 
\end{equation}
There is no easy way to compute this integral so that we have to resort to a saddle point approximation, by minimizing the expression in the exponential with respect to ${\bm u}$. After some replacements the expression to optimise reads
\begin{eqnarray}
 \mathcal{F}(\bm{u}) 
                     &=& \frac{u^2}{2} - \frac{1}{\beta} \log\left(  I_0(\beta \gamma) I_0(-\beta K_2 \frac{v_G^2}{2}) + 2 \sum\limits_{j = 1}^\infty \cos(2 j \theta_\gamma) (-1)^j I_j(\beta K_2 \frac{v_G^2}{2}) I_{2j}(\beta \gamma) \right) \; . 
\end{eqnarray}
Seeking zeros of the gradient of $\mathcal{F}$, first noticing that the $2d$ minimiser is attained for $\bm{u}$ aligned with $\bm{v_G}$, which fixes $\theta_\gamma = \mathrm{arccos}(\bm{\hat{\gamma}}\cdot\bm{\hat{v}}_G) \equiv 0 [\pi]$,  we obtain an implicit equation for the optimum $\bm{u^\star}$:
\begin{equation}
 \bm{u^\star} = \pm \mathrm{sign}(K_1) \bm{\hat{v}}_G \frac{I_1(\beta \gamma) I_0(\beta \left|K_2\right| \frac{v_G^2}{2}) + \sum\limits_{j=1}^\infty (-1)^j I_j(\beta K_2 \frac{v_G^2}{2})\left(I_{2j-1}(\beta \gamma ) + I_{2j+1}(\beta \gamma) \right) }{ I_0(\beta \gamma ) I_0(\beta \left|K_2\right| \frac{v_G^2}{2}) + 2 \sum\limits_{j = 1}^\infty (-1)^j I_j(\beta K_2 \frac{v_G^2}{2}) I_{2j}(\beta \gamma)}
 \; . 
\end{equation}
Under a critical temperature, this equation has two branches of minimizing solutions, corresponding to two minima of $\mathcal{F}$ that are not at the same height due to the presence of $K_1$.~\cite{Bore2016}
One can solve this equation numerically and compare the result to the case in which $K_2 = 0$. The results are essentially the same, except that $K_2$ makes the magnetisation more impervious to $\beta K_1 v_G$, as shown in Fig.~\ref{fig:K2m}. 
This can be explained from the equation itself, as the sums over Bessel functions are naturally perturbative because of properties of Bessel functions of growing parameter \cite{Abramowitz1972}, with larger corrections when $\beta K_1 \alpha$ increases.

\section{Changing the geometry and dimensionality\label{app:GeoDim}}

We here discuss the generality of the results described in the main text when changing the geometry of the simulation box in $2d$, or increasing the dimension to $3d$.

\subsection{Effect of the geometry}

\begin{figure}[h!]
    \centering
    \includegraphics[width= .48\columnwidth]{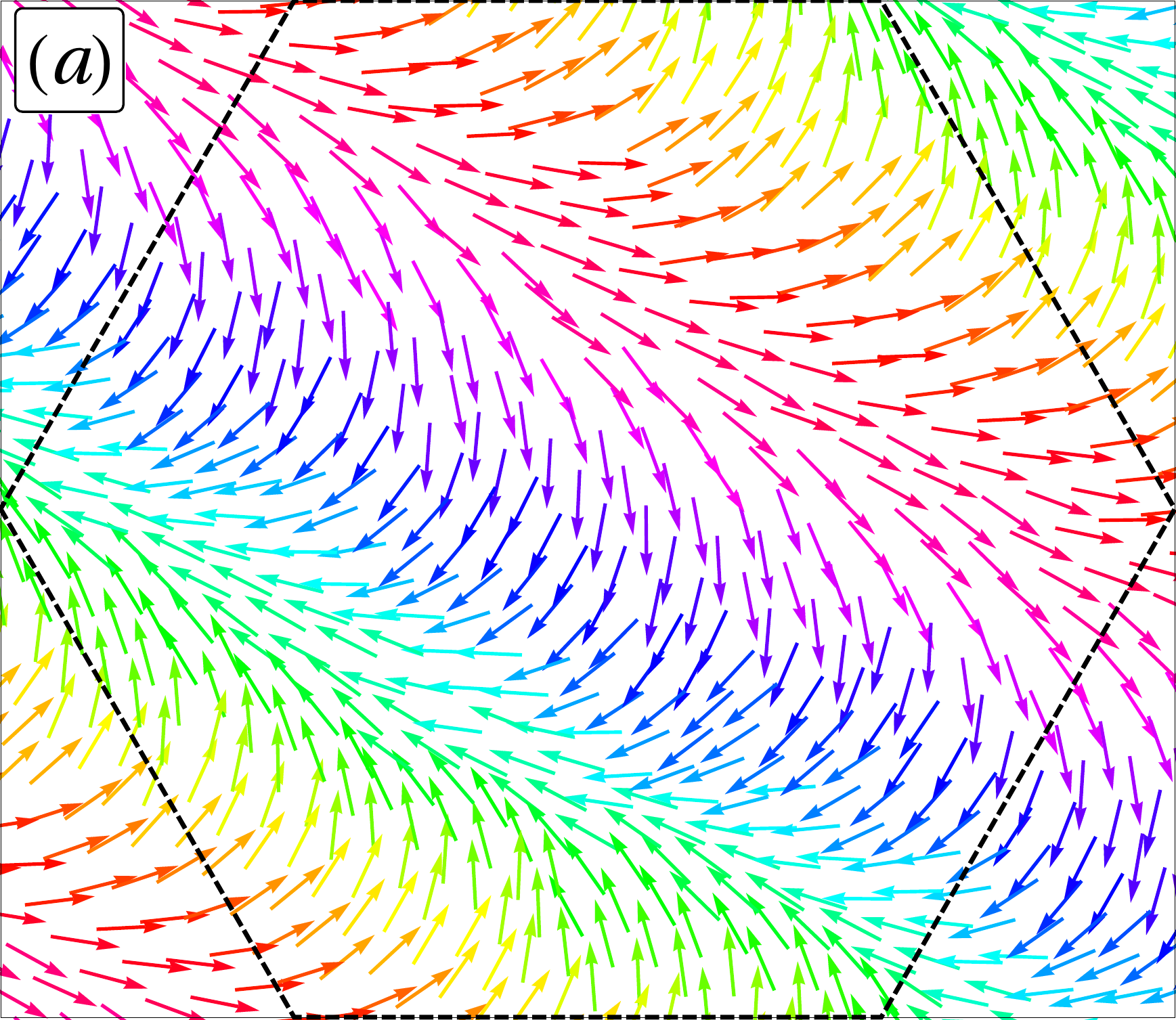}
    \includegraphics[width= .48\columnwidth]{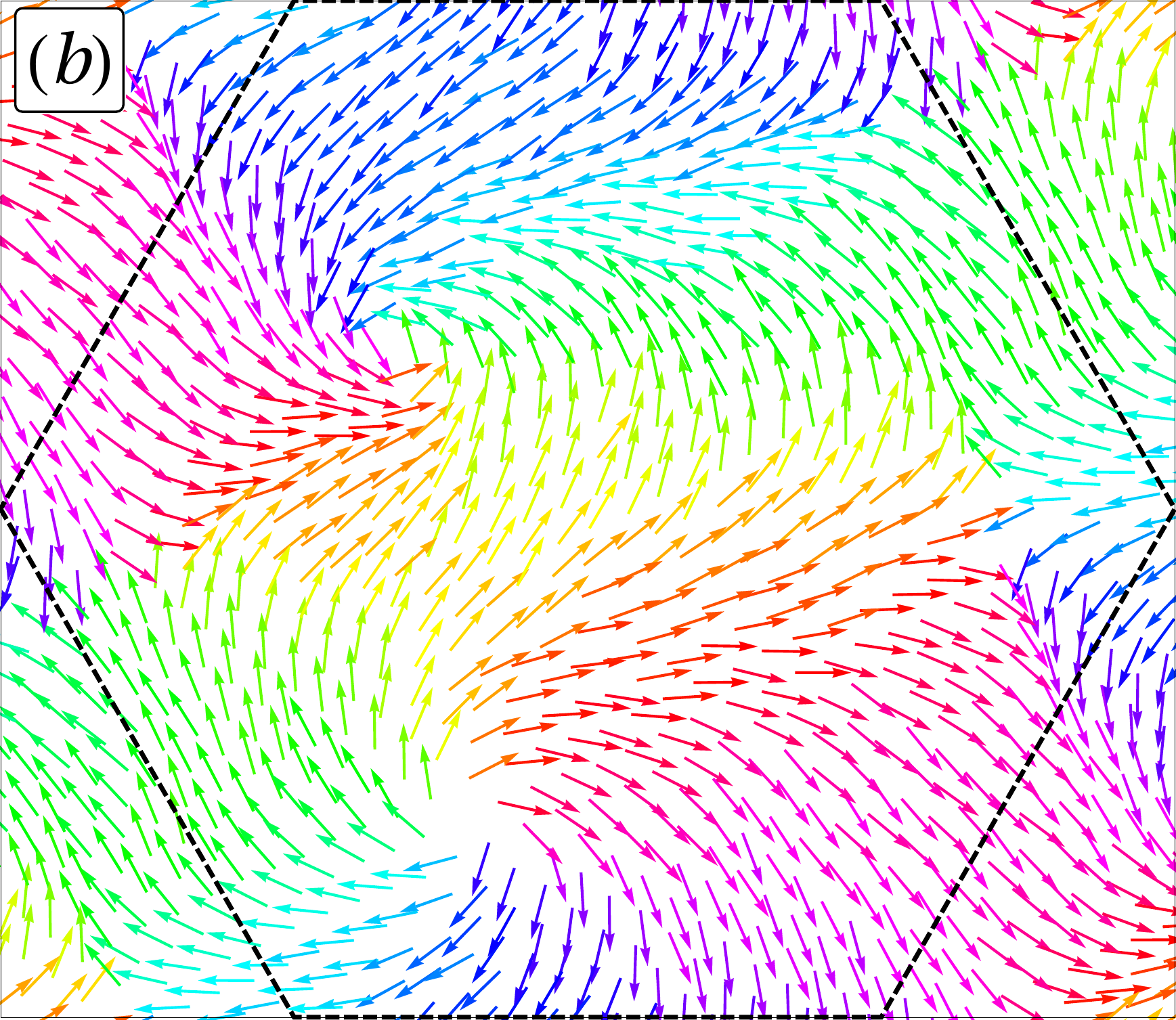} \\
    \includegraphics[width= .96\columnwidth]{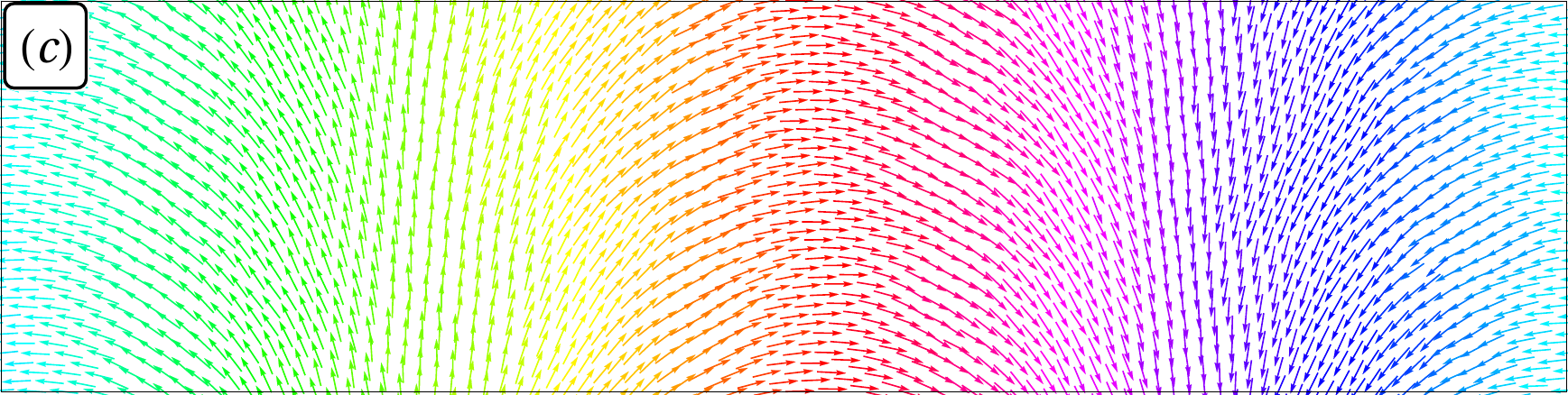}
    \caption{\textbf{Effect of Boundary Conditions.}
    Numerical ground state with hexagonal periodic boundary conditions 
    for $(a)$ $N = 512, \, K = 1.0$ and $(b)$ $N = 800, \, K = 1.0$.
    $(c)$ Numerical Ground State in a periodic rectangle for $N = 2048, \, K = 0.5$.
    In $(a)$ and $(b)$, the simulation box is showed as a dashed black line.}
    \label{fig:BCs}
\end{figure}

In the main text, every simulation was performed in a periodic square box.
Topologically, this box is homotopic to a Euclidean (flat) torus~\cite{Sausset2007}. This geometry imposes some constraints. For example,  the total number of vortices has to be compensated by the total number of antivortices.
Changing the topology of the simulation box could be interesting, as it would lead to different proportions of vortices and antivortices, and perhaps to whole new defects.
The role of geometry in finite size is less clear: one could wonder whether the configurations of 2 vortices and 2 antivortices simply come from the square shape of the box, for instance.
To address this question in a minimal way, we simulate our system in a periodic hexagon (which is also equivalent to a Euclidean torus~\cite{Sausset2007}) and in a periodic rectangle.
Examples of a few ground states obtained under these conditions are shown in Fig.~\ref{fig:BCs}.
We observe the same phenomenology in hexagonal boxes as in square ones: at low $(K,N)$, the observed non-magnetic states are solitons, and when $K$ or $N$ increase, 
2 vortices and 2 antivortices develop.
In rectangular boxes, the more slender the box becomes, the more solitons are favoured.
Indeed, using the argument given in the main text for the energy of a soliton, 
their cost decreases if the rotation is made smoother.
Therefore, in rectangular boxes, the solitons will select the longest side and be favoured with respect to vortex configurations.

In summary, as long as the homotopy class of the simulation box does not change, and using the few 
examples shown in Fig.~\ref{fig:BCs}, we expect the same phenomenology as in square boxes. Possibly 
there will be a shift of the crossover lines in the $(K,N)$ plane  
that can be expected to depend on the precise geometry.

\subsection{Effect of dimensionality}

In this paper, we mostly worked with the model defined on a $2d$ space and with, 
consequently, $2d$ continuous spins.
However, the crossover between magnetic and non-magnetic ground states, that is only due to an energetic trade-off, 
should survive in higher dimensionality of space and with spins of higher dimension.
In order to check that dimensionality does not play an important role with respect to this 
qualitative feature, we now focus on the case $d = 3$.

\begin{figure}[h!]
\vspace{0.25cm}
 \centering
\includegraphics[width=.5\textwidth]{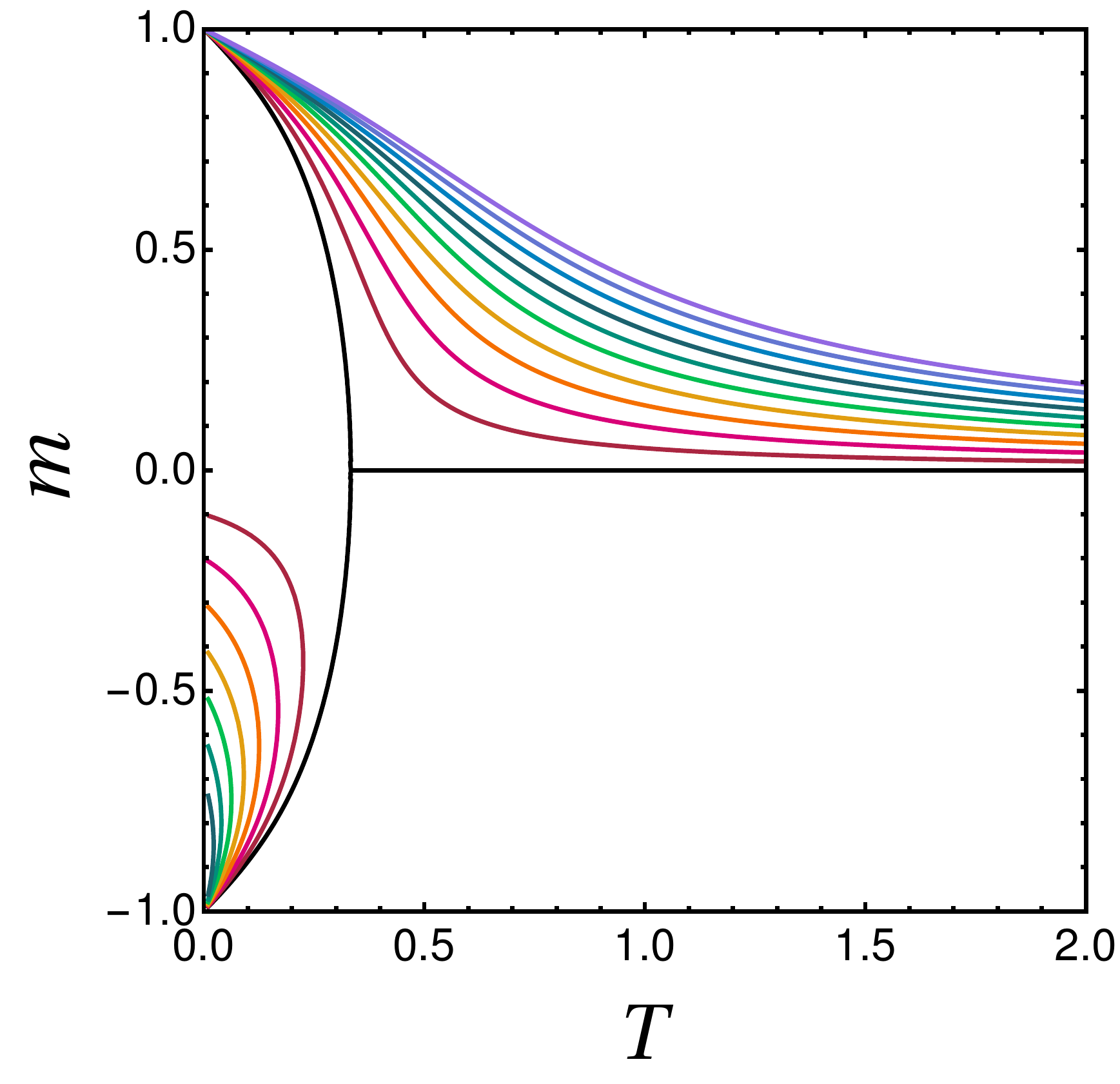}  
\caption{\textbf{Mean-field magnetisation versus temperature in $d = 3$.} All the curves shown were obtained with $K_1 = 1$, for growing values of $\alpha$ between $0$ and $1$ by steps of $0.1$. A ferromagnetic state is still observed, but the transition occurs at a temperature lower than $0.5$ since we did not rescale the interactions when going from 2 to 3 dimensions.  
\label{fig:3dsol}}
\end{figure}

First, let us revisit the mean-field calculation in $3d$ and with Heisenberg spins.
The main difference with respect to the $2d$ case with planar spins comes from the angular dependence in the momentum associated to the spins. More precisely, 
the Hamiltonian we want to study now is
\begin{eqnarray}
 H \equiv 
 \sum\limits_{i=1}^N\left\{ \frac{\bm{{p_i}}^2}{2} + 
 \frac{p_{\theta_i}^2}{2} + \frac{p_{\phi_i}^2}{2 \sin^2\theta_i} 
 -  K_1 \, \bm{p_i}\cdot\bm{s_i} - \frac{1}{2}\sum\limits_{k(\neq i)} \left[J(r_{ik})\bm{s_i}\cdot\bm{s_k} - u(r_{ik})\right] \right\}
 \; . 
\end{eqnarray}
Let us focus on the angular part of the partition function: the $p_{\theta_i}$ contribution 
is simply a Gaussian integral that gives a constant. However, the integral over $p_{\phi_i}$ looks a bit more dangerous for future calculations because of the $\sin\theta_i$ factor. The first (Gaussian) integral reads:
\begin{equation}
 \int\limits_{\mathbb{R}} dp_{\phi_i} \; 
 \exp\left(- \beta \frac{p_{\phi_i}^2}{2 \sin^2\theta_i} \right) = \sqrt{\frac{2\pi \sin^2\theta_i}{\beta}}
 \; . 
\end{equation}
Then, using the fact that $0\leq\theta_i\leq\pi$, one can simply take the $\sin$ out of the square root for all values of $i$. Therefore, for each particle, we simply obtain a $\sin\theta_i$ factor that should be integrated over $\theta_i$. In fact, this is simply the factor we were missing in the partition function to recover $d\bm{s_i} = \sin\theta_i d\theta_i d\phi_i$! We thus find that the spins should be integrated over a sphere, and not simply over a cube in angular coordinates. 
In fact, the same trick can be reproduced for any higher dimension. In the end, we will simply recover the hyperspherical infinitesimal surface element. Therefore, in any $d\geq 3$, we get, 
after some usual algebra, the saddle-point solution
\begin{eqnarray}
     \log Z_d    &\approx&  N \left[  -\left(d - \frac{1}{2}\right)\log\beta + \log V + \frac{\beta}{2} \alpha^2  - \beta \mathcal{F}(\bm{u^\star}) + cst   \right]
\end{eqnarray}     
where, as in the $2d$ case, we have to minimise a function of $\bm{u}$, an intermediate integration variable, defined through
\begin{eqnarray}
     && \mathcal{F}(\bm{u}) \equiv \frac{u^2}{2} - \frac{1}{\beta}\log\left( \frac{ I_{\frac{d}{2}-1}\left(\beta \gamma \right)}{\left(\beta \gamma \right)^{\frac{d}{2}-1}} \right) \qquad\qquad \mbox{with} \qquad\qquad
     \bm{\gamma} \equiv \bm{u} + K_1 \bm{\alpha} 
     \; . 
\end{eqnarray}  
The value of the minimizing value, $\bm{u^\star}$, is given by an implicit equation, and $\bm{u^\star}$ is still the value of the mean magnetisation in the saddle-point approximation. The results in any dimensionality are essentially similar, except that in the absence of a rescaling of interactions with $d$, the ferromagnetic transition that we observe is sent to lower temperatures. We show the particular $d = 3$ case in Fig.~\ref{fig:3dsol}.

\begin{figure}[h!]
    \centering
    \includegraphics[width = .24\columnwidth,height = .24\columnwidth]{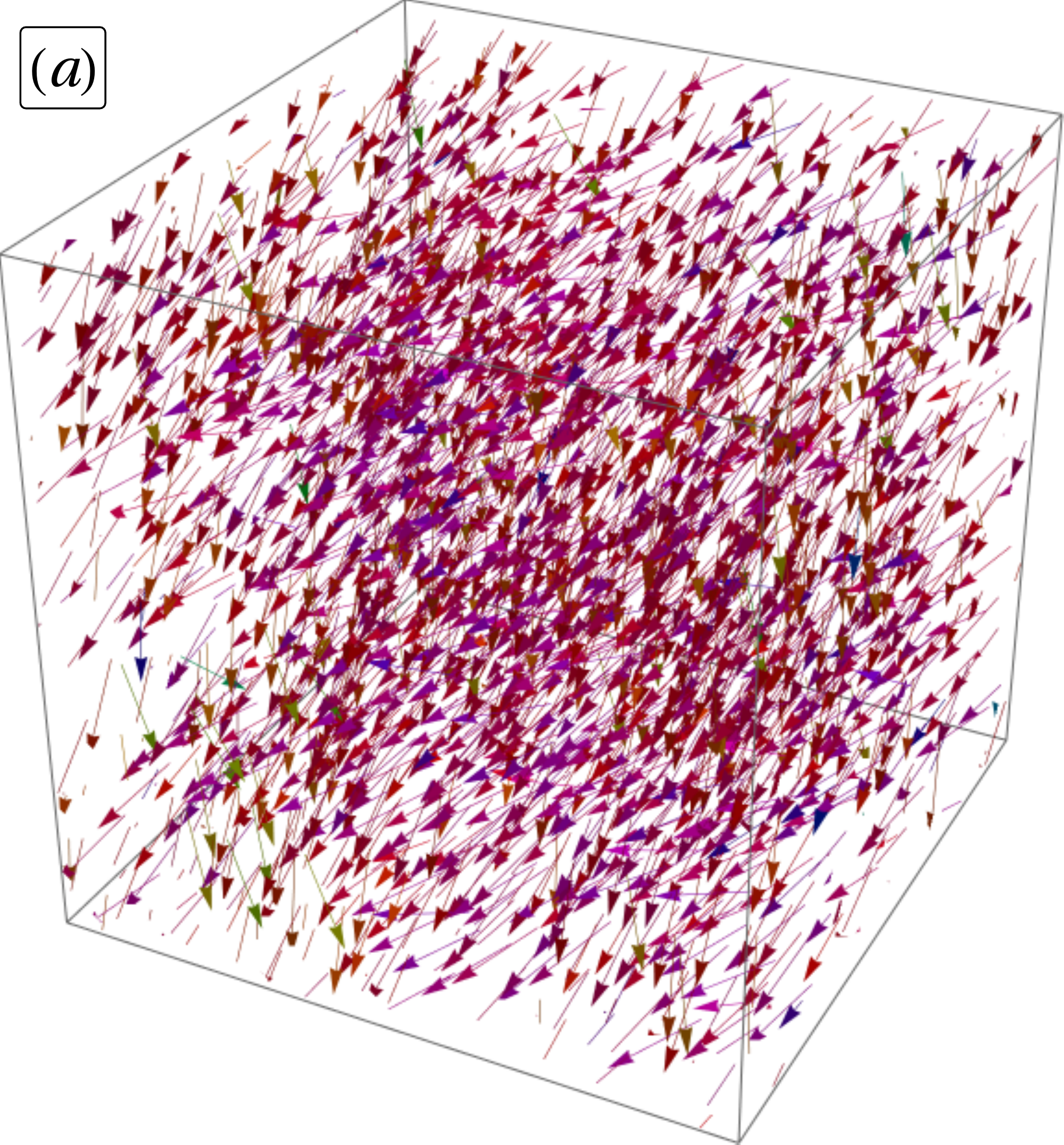}
    \includegraphics[width = .24\columnwidth,height = .24\columnwidth]{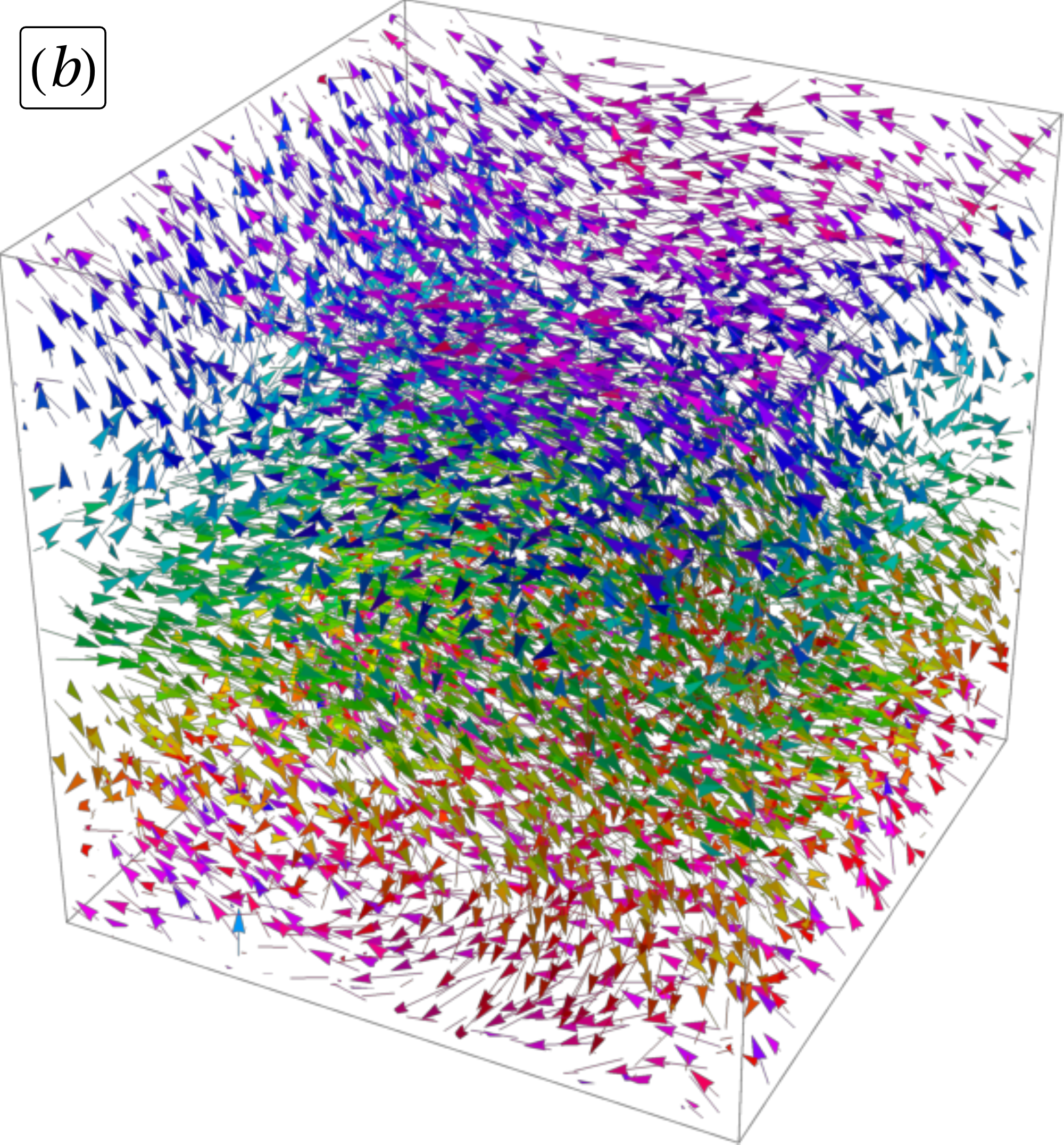}
    \includegraphics[height = .24\columnwidth]{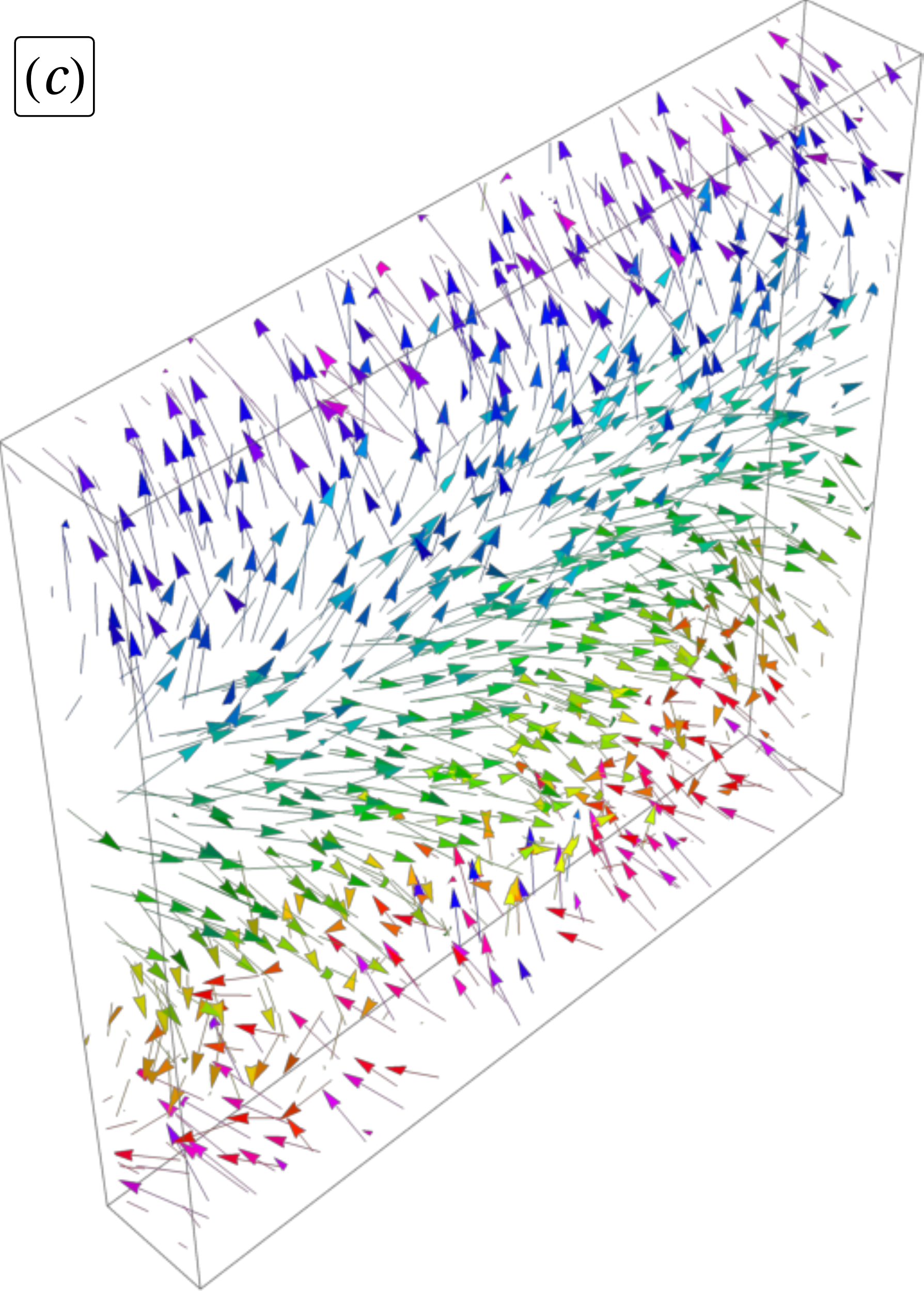}
    \includegraphics[height = .24\columnwidth]{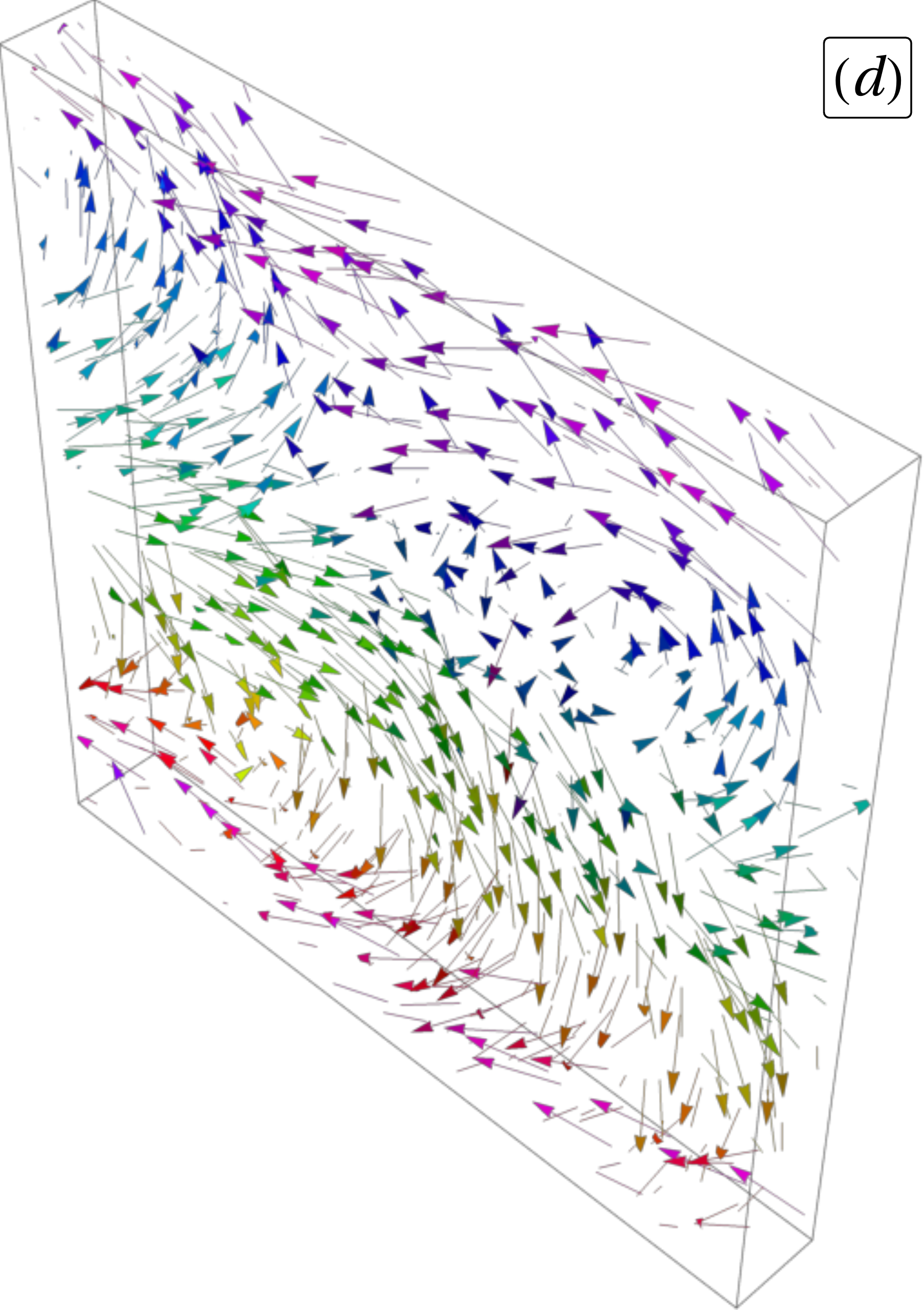}
    \caption{{\textbf{3d frustrated states.}}
    $(a)$ Low-temperature polar state in a system with $N = 2000, \, K = 0.1$.
    $(b)$ Low-temperature soliton state in a system with $N = 5000, \, K = 0.5$
    $(c)$ and $(d)$ are slices of $(b)$, that make the soliton structure more apparent.
    In the top part of $(d)$, a structure reminiscent of a vortex-antivortex pair can also be 
    seen.
    }
    \label{fig:3d}
\end{figure}

Having checked that, at the mean-field level, results for $d = 3$ are qualitatively the same as in $d = 2$, we now reproduce the same simulations as in the $2d$ case in $3d$, with classical Heisenberg spins instead of XY spins, and obtain ground states for a few values of $K$ and $N$ and a density such that no phase separation is observed.
A few snapshots thus obtained in a periodic cubic box, and for a $3d$ packing fraction $\phi_{3d} \equiv \frac{N \pi \sigma^3}{48 L^3} \approx 0.32$, are shown in Fig.\ref{fig:3d}.
and confirm that polar states and solitons are still observed in that case when varying $K$ and $N$.
In Fig.~\ref{fig:3d}$(d)$, which shows a slice of the soliton state shown in Fig.~\ref{fig:3d}$(b)$, a $2d$-like vortex-antivortex structure is also seen.
A priori, we could expect true $3d$ point defects to arise in $3d$ as well for larger systems and, possibly, larger values of $K$.

\end{document}